\documentclass[%
 reprint,
 onecolumn,
 amsmath,amssymb,
 aps,
 superscriptaddress,
]{revtex4-2}

\usepackage{graphicx}
\usepackage{dcolumn}
\usepackage{bm}
\usepackage{color}

\begin{document}


\title{Visible nonlinear photonics via high-order-mode dispersion engineering}

\author{Yun Zhao}
 \email{yz3019@columbia.edu}
 \affiliation{Department of Electrical Engineering, Columbia University, New York, NY 10027, USA}
\author{Xingchen Ji}
 \affiliation{Department of Electrical Engineering, Columbia University, New York, NY 10027, USA}
 \affiliation{School of Electrical and Computer Engineering, Cornell University, Ithaca, NY 14853, USA}
\author{Bok Young Kim}
 \affiliation{Department of Applied Physics and Applied Mathematics, Columbia University, New York, NY 10027, USA}
\author{Prathamesh S. Donvalkar}
 \affiliation{Department of Applied Physics and Applied Mathematics, Columbia University, New York, NY 10027, USA}
 \affiliation{School of Applied and Engineering Physics, Cornell University, Ithaca, NY 14853, USA}
\author{Jae K. Jang}
 \affiliation{Department of Applied Physics and Applied Mathematics, Columbia University, New York, NY 10027, USA}
\author{Chaitanya Joshi}
 \affiliation{Department of Applied Physics and Applied Mathematics, Columbia University, New York, NY 10027, USA}
 \affiliation{School of Applied and Engineering Physics, Cornell University, Ithaca, NY 14853, USA}
\author{Mengjie Yu}
 \affiliation{School of Electrical and Computer Engineering, Cornell University, Ithaca, NY 14853, USA}
 \affiliation{Department of Applied Physics and Applied Mathematics, Columbia University, New York, NY 10027, USA}
 \author{Chaitali Joshi}
 \affiliation{Department of Applied Physics and Applied Mathematics, Columbia University, New York, NY 10027, USA}
 \affiliation{School of Applied and Engineering Physics, Cornell University, Ithaca, NY 14853, USA}
\author{Renato R. Domeneguetti}
 \affiliation{Instituto de Fisica, Universidade de São Paulo, P.O. Box 66318, São Paulo 05315-970, Brazil}
\author{Felippe A.S. Barbosa}
 \affiliation{Department of Electrical Engineering, Columbia University, New York, NY 10027, USA}
\author{Paulo Nussenzveig}
 \affiliation{Instituto de Fisica, Universidade de São Paulo, P.O. Box 66318, São Paulo 05315-970, Brazil}
\author{Yoshitomo Okawachi}
 \affiliation{Department of Applied Physics and Applied Mathematics, Columbia University, New York, NY 10027, USA}
\author{Michal Lipson}
 \affiliation{Department of Electrical Engineering, Columbia University, New York, NY 10027, USA}
\author{Alexander L. Gaeta}
 \affiliation{Department of Applied Physics and Applied Mathematics, Columbia University, New York, NY 10027, USA}

\date{\today}

\begin{abstract}
Over the past decade, remarkable advances have been realized in chip-based nonlinear photonic devices for classical and quantum applications in the near- and mid-infrared regimes. However, few demonstrations have been realized in the visible and near-visible regimes, primarily due to the large normal material group-velocity dispersion (GVD) that makes it challenging to phase match third-order parametric processes. In this paper, we show that exploiting dispersion engineering of higher-order waveguide modes provides waveguide dispersion that allows for small or anomalous GVD in the visible and near-visible regimes and phase matching of four-wave mixing processes. We illustrate the power of this concept by demonstrating in silicon nitride microresonators a near-visible modelocked Kerr frequency comb and a narrow-band photon-pair source compatible with Rb transitions. These realizations extend applications of nonlinear photonics towards the visible and near-visible regimes for applications in time and frequency metrology, spectral calibration, quantum information, and biomedical applications.
\end{abstract}

\maketitle

\section{Introduction}

With the rapid development of nanofabrication techniques, nonlinear- and quantum-based applications are being realized in chip-scale devices. In order to operate with high efficiency, third-order parametric nonlinear processes must satisfy phase matching (PM) conditions largely governed by the group-velocity dispersion (GVD). In photonic waveguides the GVD has contributions from the material and from the waveguide confinement. The waveguide GVD can be tuned by changing the structure dimensions \cite{Turner_OE_2006}, which for four-wave mixing requires anomalous or near-zero GVD. This is critical for applications such as  Kerr frequency comb generation (KCG) \cite{Kippenberg_Science_2018,Gaeta_NatPhot_2019} and photon-pair generation (PPG) \cite{Kues_NatPhot_2019}. For all photonic materials [e.g., silicon nitride (SiN), silica, \textit{etc.}] the GVD becomes highly normal at shorter wavelengths, which makes it impossible to satisfy the PM conditions with conventional waveguide dispersion engineering. Overcoming this obstacle is essential to developing chip-based photonics in the visible and near-visible regimes.

\begin{figure*}
	\centering
	\includegraphics{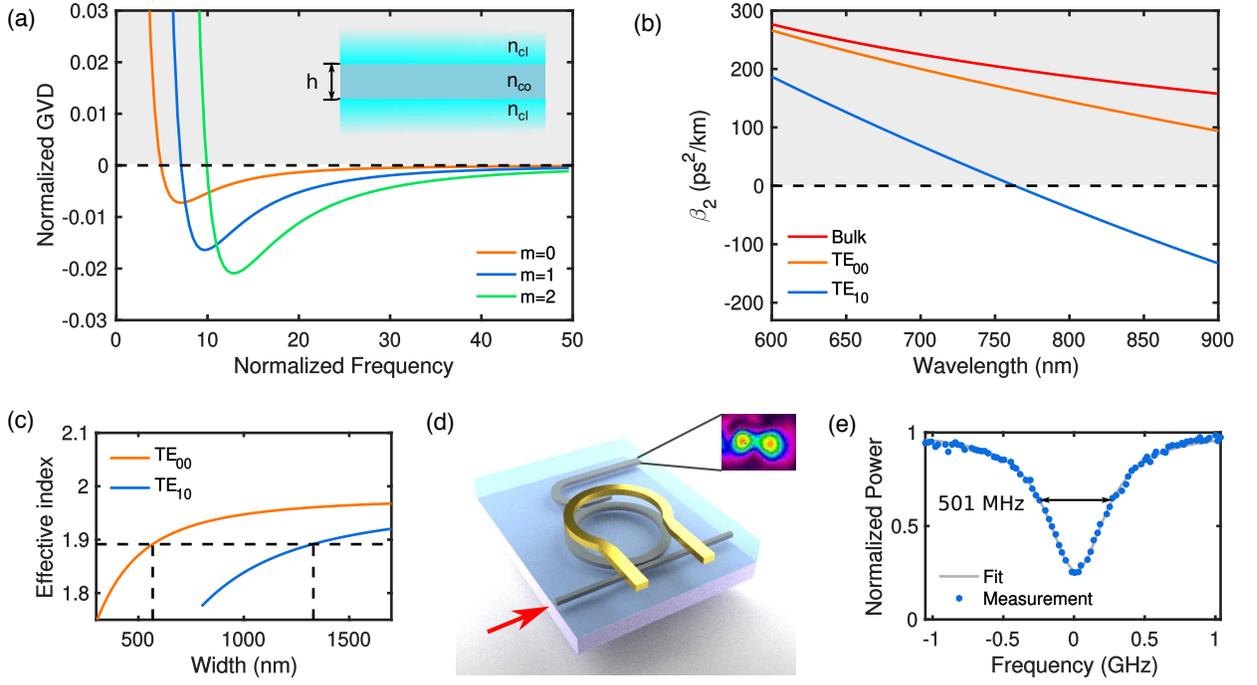}
	\caption{(a) Normalized GVD. Inset, structure of the slab waveguide. (b) Simulated GVD for a 730 $\times$ 1330 nm microring resonator with a bending radius of 22 $\mu$m. (c) Effective refractive index of the TE$_{00}$ mode of a straight waveguide and TE$_{10}$ mode of a ring. Efficient mode conversion is realized when the two modes have the same effective indices. (d) Device schematic. Inset, measured mode profile. (e) Measured resonance.}
	\label{figDesign}
\end{figure*}

In this paper, we perform dispersion engineering by utilizing the high-order-modes of chip-based waveguides to create anomalous GVD across a large range of wavelengths in the near-visible and visible regimes. We demonstrate the power of this dispersion engineering by generating broadband Kerr combs near 784 nm, which represents the first soliton comb at this wavelength regime using a monolithically integrated chip-based platform. The low wavelength side of the comb reaches visible regime (\textless 740 nm, the definition of visible regime may vary), which are the shortest wavelength components generated directly by a Kerr frequency comb. We also show that this dispersion engineering enables a silicon chip-based narrow-band photon-pair source at near-visible wavelengths and is compatible with rubidium quantum memories (operating near 795 nm). We also show through simulation that combs deep in the visible are possible, which overlap Hg$^+$ and Yb transitions.

\section{Dispersion engineering using high-order modes}

The refractive indices of most materials can be modeled by a simple two-level system \cite{Boyd_NO}, which yields the Sellmeier equation. Such a model yields a refractive index $n(\omega)$ that scales as $(\omega_0^2-\omega^2)^{-1/2}$, where $\omega_0$ is the frequency of the material resonance. For visible light, the dominating resonance frequency corresponds to the bandgap energy, which creates strong normal GVD as the light frequency shifts towards to it.

To illustrate the power of dispersion engineering using high-order waveguide modes, we start with normalized parameters \cite{Pollock_IP} for a step-index slab waveguide [inset, Fig. \ref{figDesign}(a)]. Our results capture the behavior of rectangular waveguides since their effective refractive index can be approximated by two slab waveguides with the effective index method \cite{Pollock_IP}. The two essential parameters are,
\begin{align}
&V = \frac{\omega}{c} h \sqrt{n_{co}^2-n_{cl}^2} \label{eqNormV},\\
&b \approx \frac{2n_{co}n_{eff}}{n_{co}^2-n_{cl}^2},
\end{align}
where $V, b$ are the normalized frequency and refractive index, respectively, $c$ is the speed of light, $h$ is the waveguide height, $n_{co}$ and $n_{cl}$ are the refractive indices of the core and the cladding, respectively. For TE modes, $V$ and $b$ obey the transcendental equation,
\begin{align}\label{eqEigen}
V\sqrt{1-b} = m\pi + 2\text{tan}^{-1}\sqrt{b/(1-b)},
\end{align}
where $m$ is the mode order. Equation (\ref{eqEigen}) is useful in analyzing the dispersion properties of waveguides \cite{Bennett_AO_1980}, and we focus our attention on the waveguide dispersion by setting the material dispersion to zero, that is, $n_{co}$ and $n_{cl}$ are independent of $\omega$. The waveguide GVD can then be written as,
\begin{align}\label{eqbeta2}
\beta_{2,wg} = (n_{co}^2-n_{cl}^2)^{\frac{3}{2}}\frac{h}{2n_{co}c^2}B_2(V),
\end{align}
where $B_2(V) = \frac{d^2(Vb)}{dV^2}$ is the normalized GVD and contains the general waveguide GVD properties. Due to the transcendental nature of Eqn. (\ref{eqEigen}), $B_2$ is calculated numerically. From Fig. \ref{figDesign}(a), we observe that close to the cutoff frequency, $B_2$ is strongly positive (normal). As the frequency increases, $B_2$ becomes strongly negative (anomalous) and then gradually decreases in magnitude while maintaining its sign. More importantly, we find that the higher-order modes can produce larger negative $B_2$ than the lower-order ones. A similar idea is shown for whispering gallery mode resonators (WGMR) where the analysis takes a different approach \cite{Savchenkov_NatPhot_2011}. 

Designing a waveguide that compensates for the large normal GVD of photonic materials in the near-visible and visible regimes requires access to the large negative part of $B_2$. This can be achieved by using high-order modes and small waveguide structures. In this work, we base our devices on the SiN platform. We use a finite-element method (FEM) mode solver with a well characterized Sellmeier equation for our SiN thin film \cite{Luke_OL_2015} to accurately model the total GVD of rectangular waveguides and microresonators. Although stronger anomalous GVD can be realized with a smaller waveguide size, the reduction in size also introduces additional losses due to an increase in scattering at the core-cladding interface. For the current experiments, we choose a waveguide cross section of 730 nm $\times$ 1330 nm and a ring resonator radius of 22 $\mu$m, with a width fabrication uncertainty of $\pm$ 40 nm. The simulated GVD is plotted in Fig. \ref{figDesign}(b). We operate at the TE$_{10}$ mode whose zero-GVD point is near 760 nm. We design the bus-ring coupling region so that the TE$_{00}$ mode of the bus excites the TE$_{10}$ mode of the ring \cite{Luo_NatComm_2014}. This is done by matching the effective refractive index of the two modes, which leads to bus dimensions of 730 $\times$ 568 nm [Fig. \ref{figDesign}(c)]. In this way, we ensure that the input and output modes are fundamental even though the resonating modes are in higher-order, making this device easy to interface with any other on-chip component. To verify that the TE$_{10}$ mode is indeed excited, in selected devices we include a drop port, that has the same cross section as the ring. The mode profile in the ring can be imaged at the output of the drop-port [top inset, Fig. \ref{figDesign}(d)] with a microscope configuration and a camera. These drop-port devices, however, are not used for KCG or PPG since the extra coupling loss reduces the \textit{Q}. The final schematic of the device is shown in Fig. \ref{figDesign}(d). We characterize the \textit{Q} of our devices using a wavemeter-calibrated frequency scan and a Lorentzian fit. The full width at half maximum (FWHM) is measured to be 501 MHz, which corresponds to a loaded Q of 7.7 $\times$ 10$^5$ with an extinction ratio of 74\%.

\section{Kerr comb generation}

\begin{figure*}
	\centering
	\includegraphics{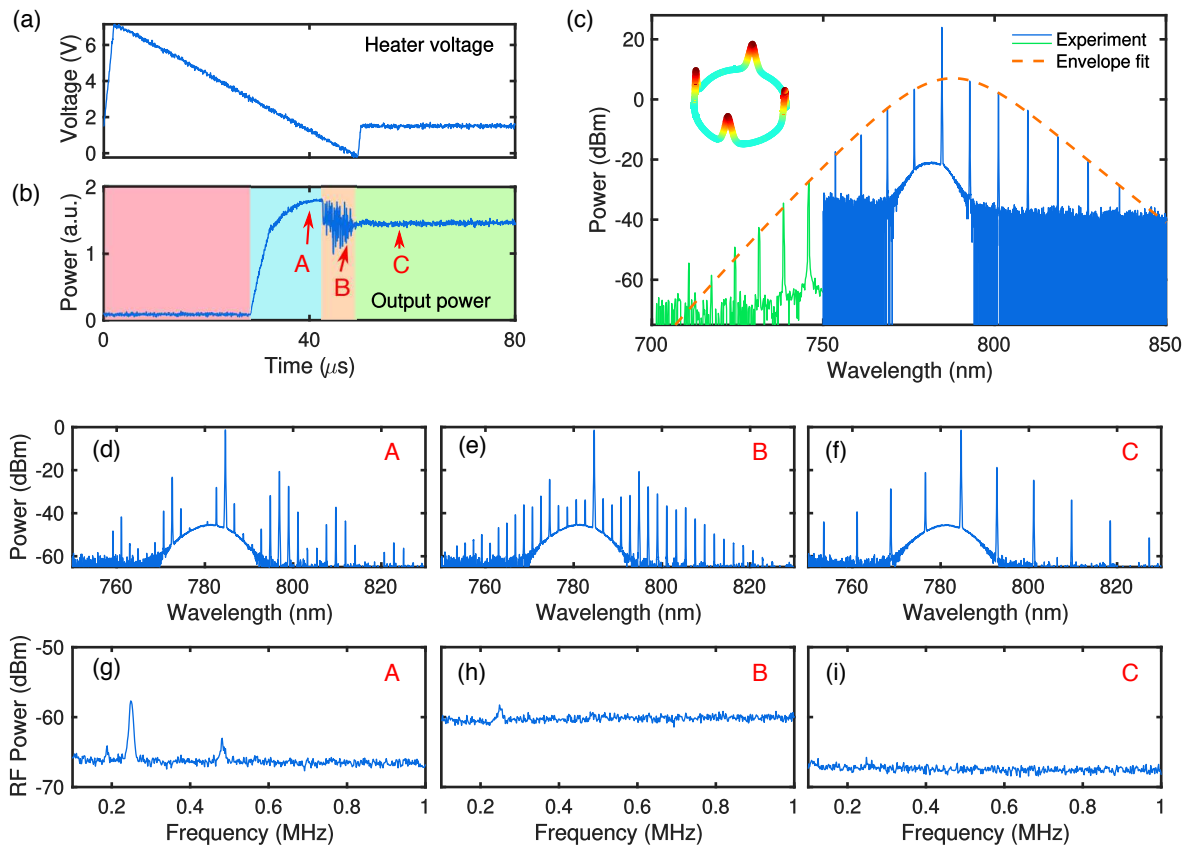}
	\caption{(a) Voltage pattern used to modulate the on-chip heater. (b) The evolution of comb power corresponding to the pattern in (a). (c) Experimental comb spectrum calibrated for collection losses. Inset, simulated intracavity 4-soliton state. (d, e, f) The optical spectra corresponding to point A (minicomb formation), B (chaotic state) and C (soliton state) in (b), respectively. (g, h, i) The RF noise corresponding to (d) - (f).}
	\label{figSpectrum}
\end{figure*}

The first application we demonstrate using this high-order-mode nonlinear microresonator is Kerr frequency comb generation reaching the visible regime. Kerr frequency combs provide a pathway towards fully integrated comb sources that could find applications across numerous areas of science and engineering \cite{Herr_NatPhot_2014, Stern_Nature_2018, Kippenberg_Science_2018,Gaeta_NatPhot_2019}, including spectroscopy \cite{Suh_Science_2016, Pavlov_OL_2017, Yu_NatComm_2018, Dutt_SciAdv_2018}, ranging \cite{Trocha_Science_2018, Suh_Science_2018}, frequency synthesizer \cite{Spencer_Nature_2018}, and coherent communication \cite{MarinPalomo_Nature_2017}. Moreover, extending KCG to near-visible and visible wavelengths enables additional applications in optical clockwork \cite{Vanier_ApplPhysB_2005, Papp_Optica_2014}, astronomical spectrograph calibration \cite{Li_Nature_2008, Obrzud_NatPhot_2019, Suh_NatPhot_2019}, and biological imaging \cite{Lee_Misc_2001, Fercher_ReportsProgressPhys_2003}. Additionally, these wavelength regimes are compatible with the low-cost high-quality Si-based detectors and cameras. Various schemes have been proposed and tested in the past, including direct generation \cite{Saha_OE_2012, Yang_OL_2016, Wang_LaserPhotRev_2016, Soltani_LaserPhotRev_2016} and frequency conversion from telecom combs \cite{Miller_OE_2014, Guo_PRAppl_2018}, but to date, modelocking has only been demonstrated by Lee, \textit{et al.} \cite{Lee_NatComm_2017} near 778 nm using a silica WGMR. While SiN has proven to be highly promising platform in the near-IR, at near-visible wavelengths, SiN has a material GVD that is more than 5 times higher than that of silica used in \cite{Lee_NatComm_2017}. To date, the lowest wavelength achieved by a modelocked Kerr comb on this platform is demonstrated by Yu, \textit{et al.} near 770 nm \cite{Yu_PRAppl_2019}, by generating a dispersive wave fed by a main soliton pumped at 1 $\mu$m. The dispersive wave process is governed by the higher-order dispersion coefficients, making it highly sensitive to fabrication variations \cite{Pfeiffer_Optica_2017}. It is thus advantageous to generate a soliton centered in this wavelength regime since it also can provide higher comb power as compared to dispersive waves.

We generate a soliton-modelocked comb using the device described in the previous section with the thermal tuning method as demonstrated by Joshi, \textit{et al}. \cite{Joshi_OL_2016}. Specifically, we fix the pump laser frequency to the blue side of the resonance and decrease the heater voltage which blue-detunes the resonance. At the output, we record the comb spectrum and power. The output power curve [Fig. \ref{figSpectrum}(b)] is composed of four segments that correspond to four different dynamical processes leading to the final modelocked comb, namely subthreshold state, Turing pattern state, chaotic state and soliton state. The heater voltage is slightly increased after the transition into soliton state (known as the ``soliton step'') to compensate for the temperature difference owing to an abrupt change of intracavity power. In Fig. \ref{figSpectrum}(c), we show the spectrum of a 4-soliton state where the solitons are equally spaced inside the ring [inset, Fig. \ref{figSpectrum}(c)]. This is the preferred state since it has an intracavity power that is close to that of the chaotic state, making it experience the least amount of thermal backlash after the ``soliton step", which we find to be more pronounced than in the near IR. For pumping at 782 nm, which is one free spectral range (FSR) away from 784 nm, the 5-soliton state is found to be most stable, indicating that a mode interaction near 792 nm is the dominant factor in determining the pattern of the soliton state \cite{cole_NatPhot_2017}. Limited by the dynamic range of the OSA ($\approx$ 60 dB), we obtain the spectrum in two shots under the same experimental conditions. We observe comb lines down to 710 nm, which represents the shortest wavelength generated by a modelocked Kerr comb. We then fit our experiment spectrum with a sech$^2$ envelope which yields a 3-dB bandwidth of 9.9 THz.

We further characterize the noise properties of the generated soliton state with a radio frequency spectrum analyzer (RFSA). We gradually tune the pump laser into resonance and record the noise at three different stages of KCG process [indicated in Fig. \ref{figSpectrum}(b)], namely the minicomb state (A), chaotic state (B) and soliton state (C). As observed previously \cite{Joshi_OL_2016}, the minicomb stage has a low noise floor with sharp peaks, the chaotic state has broadband and high noise, and the soliton state has a flat noise spectrum and the lowest RF noise of all three stages. 

\begin{figure*}
	\centering
	\includegraphics{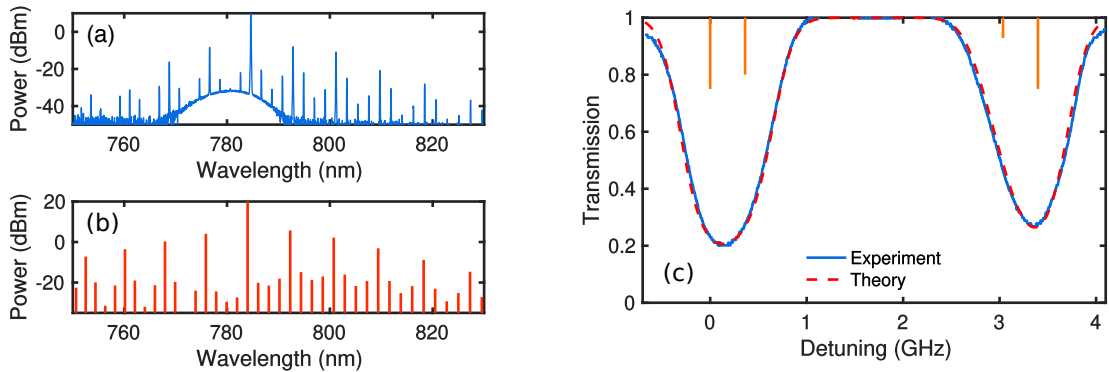}
	\caption{(a) Asymmetric 4-soliton state with denser comb line spacing. (b) Simulated spectrum of (a). (c) Rb D1 transition probed by a comb line of our near-visible comb. The Doppler free transitions and their relative strengths are shown as vertical sticks.}
	\label{figCombScan}
\end{figure*}

One appealing property of the generated near-visible combs is that they span several atomic transitions that are commonly used for time and frequency metrology. For example, the transitions allow for full stabilization of the comb by locking two of the comb lines to two atomic transitions \cite{Papp_Optica_2014}. Here, we show the interaction between our comb and rubidium (Rb) atoms by scanning a comb line near 795 nm across the well known Rb D1 transition. To generate such a comb line from our 4-soliton state, we increase our pump power from 220 mW to 270 mW. At large pump-cavity detuning, we observe weaker comb lines appearing between the strong comb lines [Fig. \ref{figCombScan}(a)]. This is caused by the disruption of the perfect symmetric soliton locations due to high power, which is well modeled in our simulations [Fig. \ref{figCombScan}(b)]. We use a bandpass filter to select the comb line near 795 nm and send it through a 2-cm-long $^{85}$Rb pure isotope cell held at 53.7 $^\circ$C. As a result of soft thermal locking \cite{Carmon_OE_2004}, this comb line can be tuned by up to 10 GHz by simply tuning the pump wavelength. In our experiment, we tune the pump laser by 5 GHz through piezoelectric tuning and record the transmission of the 795-nm comb line [Fig. \ref{figCombScan}(c)]. We also fit our experimental data to the theoretical calculations based on published properties of Rb \cite{Siddons_JPhys_2008}. The fitting parameters are the atomic density, the location of the zero detuning (defined by the F = 3 to F$^\prime$ = 2 transition), and the scanning range. We observe good agreement between theory and experiment, indicating that the comb lines are strongly interacting with atoms. The slight lower transmission at the two extremes of detuning is caused by residual $^{87}$Rb atoms in the $^{85}$Rb cell.

\section{Photon-pair generation}

\begin{figure*}
	\centering
	\includegraphics{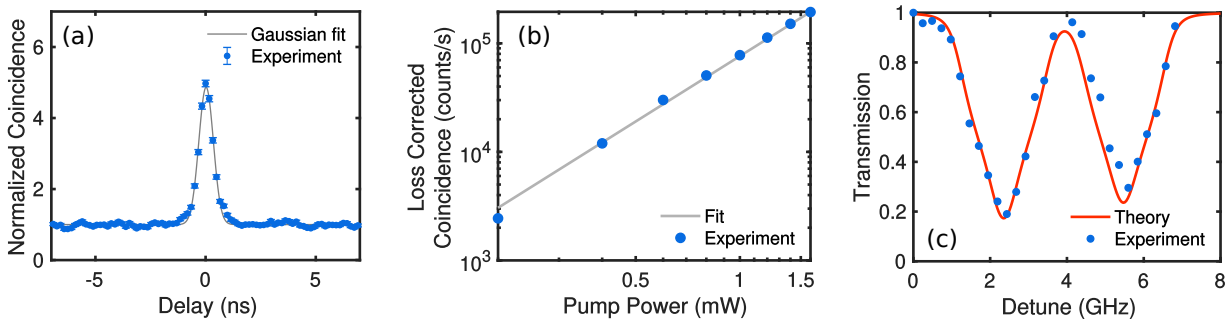}
	\caption{ \textbf(a) Normalized coincidence between signal and idler $g^{(2)}_{si}(\tau)$. (b) Photon-pair generation rate scaling. The plot is corrected for collection and detector losses. (c) Photon interaction with Rb D1 transition.}
	\label{figPhotons}
\end{figure*}

Nonlinear photonic devices in the near-visible and visible wavelength regime are important for quantum applications. First, nonclassical states can only be generated through nonlinear processes. Secondly, the existence of high quality avalanche photodetectors (APD) and well studied atomic transitions at this wavelength regime makes it extremely attractive. As an example of near-visible quantum optics in our high-order-mode nonlinear microresonator, we show correlated narrow-band photon-pair generation in this wavelength regime, which has important application in rubidium or cesium memory-based quantum information networks \cite{Liu_Nature_2001, Phillips_PRL_2001, Chaneliere_Nature_2005}. Previously, such photon-pairs have been generated via spontaneous four-wave mixing (SFWM) in warm alkali vapor \cite{Balic_PRL_2005} or via spontaneous parametric down conversion (SPDC) in a bulk lithium niobate WGMR \cite{Fortsch_NatComm_2013} and a lithium niobate Fabry-P\'erot cavity \cite{Luo_NewJPhys_2015}. PPG in SiN microresonators is promising for quantum information applications due to its scalability and low loss \cite{Ramelow_ArXiv_2015}. Moreover, the flexibility in nanofabrication provides ways for better controlling of the photon states \cite{He_Optica_2015, Vernon_OL_2016, Christensen_OL_2018}. Lu, \textit{et al.} \cite{Lu_NatPhys_2019} demonstrated the correlation between one visible and one telecom photon generated from a SiN microresonator \cite{Yu_PRAppl_2019}. However, for applications that do not require long distance fiber communication, such as free space quantum communication and memory assisted quantum computing \cite{Kok_RMP_2007}, it is desirable to generate both photons at near-visible wavelengths which has lower design and fabrication complexity. Moreover, due to their similarity, both photons can be further manipulated with same devices and be detected with room-temperature high-efficiency silicon-based APD's.

\begin{figure*}
	\centering
	\includegraphics{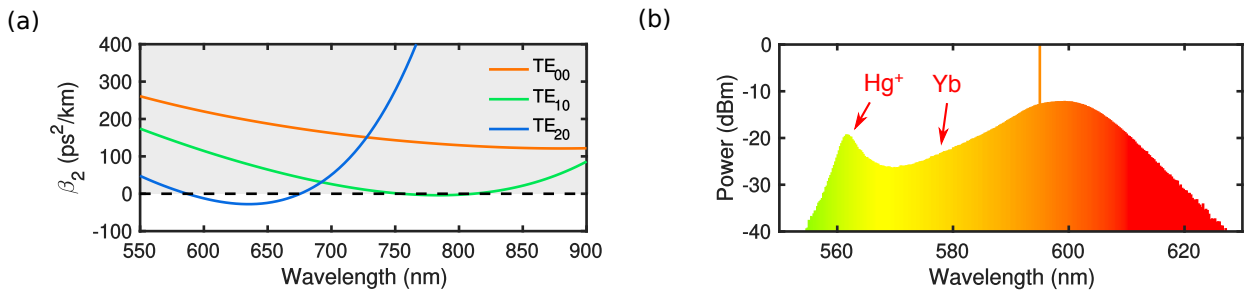}
	\caption{(a) Simulated dispersion of a microresonantor with 400 nm $\times$ 1000 nm cross section and 100 $\mu$m bending radius. (b) Simulated Kerr comb spectrum for the TE$_{20}$ mode in (a).}
	\label{figFuture}
\end{figure*}

We generate near-visible photon-pairs by driving our microresonator at pump powers below the parametric oscillation threshold. We set our pump at 784.7 nm and use bandpass filters to select photons at 794.8 nm (signal) and 774.9 nm (idler). We verify the photon correlation by measuring the coincidence function $g^{(2)}_{si}(\tau)$ between the signal and idler. We observe a clear coincidence peak at zero relative delay [Fig. \ref{figPhotons}(a)]. We also observe a significant amount of uncorrelated noise photons which is proportional to the pump power in the bus waveguide (supplementary material). This is due to SiN used in our samples being slightly Si-rich, which generates broadband fluorescence when pumped below 1.1 $\mu$m. This can be overcome by refining the fabrication process to reduce the Si concentration \cite{Smith_arXiv_2019}. Since the FWHM of each resonance is 501 MHz, the theoretical FWHM of $g^{(2)}_{si}$ is 238 ps \cite{Ou_PRL_1999}, which is comparable to the timing jitter of our single-photon counting module (SPCM), and thus cannot be readily resolved in our $g^{(2)}_{si}(\tau)$ measurement. We further characterize the generation rate by measuring the coincidence at different pump power levels. The generation rate is proportional to the square of the pump power as all data points fall on a straight line with a slope of 2 in the log-log plot [Fig. \ref{figPhotons}(b)]. We fit the measurement and achieve a photon-pair generation efficiency of 7.6$\times10^4$ pairs s$^{-1}$ mW$^{-2}$ for these narrow-band photons, which is comparable to many other cavity enhanced PPG demonstrations \cite{Clemmen_OE_2009}.

Our photon source is also easily tunable by controlling jointly the pump wavelength and the on-chip heater together. We show the tunability by sending the signal through a Rb vapor cell and tuning it across the D1 transition. Figure \ref{figPhotons}(c) shows the normalized coincidence rate versus the pump detuning. For stable operation, we blue detune the pump to operate in the soft thermal locking regime \cite{Carmon_OE_2004}. Since this is comparable to the FWHM bandwidth of the resonance, it must be incorporated into the photon model. The photon spectrum has the shape,
\begin{align}\label{eqSpectrum}
S(\omega) \propto \frac{1}{(4\omega^2+\Delta\omega^2-\mathcal{W})^2+4\Delta\omega^2\mathcal{W}},
\end{align}
where $\omega$ is the frequency detuning, $\Delta\omega$ is the FWHM angular frequency of the resonance, $\mathcal{W}$ (can assume both signs) is the combined effect of the pump detuning, dispersion and cross phase modulation (supplementary material). The theoretical curve is calculated by first measuring the Rb atomic density of the cell separately with a laser and the photon wavelength with a wavemeter by driving the ring above threshold. The theoretical transmission versus detuning curve is a convolution of the Rb absorption spectrum and the photon spectrum in Eqn. (\ref{eqSpectrum}). We also notice that a cavity mode interaction is present at the signal resonance, which effectively increases the pump detuning (supplementary material). We leave the pump detuning as the only free parameter to fit the measurement. There are $\pm1^\circ$C temperature fluctuations and constant air turbulence near the cell due to the heating. Nonetheless, the good agreement between experiment and theory indicates that the properties of photons are well characterized.

\section{Discussion}
We can take advantage of higher-order mode dispersion engineering for KCG at even shorter wavelengths. The state-of-art modelocked Ti:sapphire oscillator has been shown to produce spectral components below 600 nm \cite{Ell_OL_2001}, enabling the interaction with more species of clock atoms. Our simulations show that this can also be achieved by higher-order mode Kerr combs with realistic parameters. Following the previous argument, a tighter mode confinement is required to push the anomalous-GVD region to shorter wavelengths. In Fig. \ref{figFuture}(a), we show FEM simulation of a 400 nm $\times$ 1000 nm SiN waveguide cladded by SiO$_2$, with a bending radius of 100 $\mu$m. The TE$_{20}$ mode has a zero-GVD point at 587 nm. We simulate the corresponding soliton spectrum using the Lugiato-Lefever equation \cite{Coen_OL_2013}, where we assume a pump of 300 mW at 595 nm and a resonance FWHM of 700 MHz [Fig. \ref{figFuture}(b)]. We see the clock transitions of Hg$^+$ (563 nm) and Yb (578 nm) atoms are covered by sufficiently strong comb lines \cite{Fortier_OL_2006}. While anomalous GVD can be achieved at even shorter wavelengths with even higher order modes, the increased cavity losses caused by the use of these modes at shorter wavelengths, as well as the onset of two-photon absorption, pose significant challenges to Kerr comb generation below 550 nm via this approach.

Even though in principle PPG can occur regardless of phase matching, in practice phase matching is required to achieve a high generation rate and to avoid spurious processes. In our model for PPG with dispersion (supplementary material), we also find that phase matching contributes to the narrowness of the photon bandwidth. Notably, a blue detuned pump balances the phase mismatch that comes from a small anomalous GVD so that both high generation rate and narrow bandwidth can be achieved in this regime. 

\section{Conclusion}
In conclusion, we have shown that dispersion engineering through high-order waveguide modes can compensate for the strong normal material GVD at near-visible and visible wavelengths to realize small or anomalous GVD that is not possible with the fundamental modes. With this approach, we designed a high-order-mode nonlinear microresonator that generated the first Kerr comb from a near-visible pump on a monolithically integrated chip-based platform. This dispersion engineering capability enables potential applications such as compact atomic clocks, astrocombs and high quality bio-imaging light sources. In addition, we used the same nonlinear microresonator to demonstrate, to the best of our knowledge, the first silicon-chip-based narrow-band near-visible photon pair source which could enable applications in memory-based quantum information networks.

\textbf{Funding.} We acknowledge support from Air Force Office of Scientific Researchm(FA9550-15-1-0303) and National Science Foundation (EFMA-1641094, PHY-1707918). 

\textbf{Acknowledgement.} The authors also thank Dr. Alessandro Farsi, Dr. Sven Ramelow and Dr. Aseema Mohanty for helpful discussions.

%

\pagebreak
\widetext
\begin{center}
	\textbf{\large Visible nonlinear photonics via high-order-mode dispersion engineering: supplementary material}
\end{center}
\setcounter{equation}{0}
\setcounter{figure}{0}
\setcounter{table}{0}
\setcounter{page}{1}
\setcounter{section}{0}
\makeatletter
\renewcommand{\theequation}{S\arabic{equation}}
\renewcommand{\thefigure}{S\arabic{figure}}
\renewcommand{\bibnumfmt}[1]{[S#1]}
\renewcommand{\citenumfont}[1]{S#1}

\section{Normalized parameters}
We rederive the relation between GVD and the normalized GVD parameter with the exact form for $b$,
\begin{align}\label{eqnb}
b = \frac{n_{eff}^2-n_{cl}^2}{n_{co}^2-n_{cl}^2},
\end{align}
where the symbols are explained in the main text. With Eqn. (1) in the main text and Eqn. (\ref{eqnb}), we get,
\begin{align}
\begin{split}
\beta &= \frac{\omega n_{eff}}{c} = \frac{V}{h}\left(b+\frac{n_{cl}^2}{n_{co}^2-n_{cl}^2}\right)^\frac{1}{2}\\
&=\frac{n_{cl}}{h\sqrt{n_{co}^2-n_{cl}^2}}V + \frac{\sqrt{n_{co}^2-n_{cl}^2}}{2hn_{cl}}Vb - \frac{1}{8h}(\frac{n_{co}^2-n_{cl}^2}{n_{cl}^2})^\frac{3}{2}Vb^2 + \cdots.
\end{split}
\end{align}
Ignoring material dispersion, we get the waveguide GVD,
\begin{align}
\beta_{2,wg} = (n_{co}^2-n_{cl}^2)^{\frac{3}{2}}\frac{h}{2n_{cl}c^2}\left(\frac{d^2(Vb)}{dV^2}-\frac{n_{co}^2-n_{cl}^2}{4n_{cl}^2}\frac{d^2(Vb^2)}{dV^2}+\cdots \right).
\end{align}
For our SiN platform, $n_{co} \approx 2.0$, $n_{cl} \approx 1.5$, the second term is 5 times smaller than the first thus it can be ignored for qualitative analysis. 

\begin{figure*}[htbp]
	\centering
	\includegraphics[width=\linewidth]{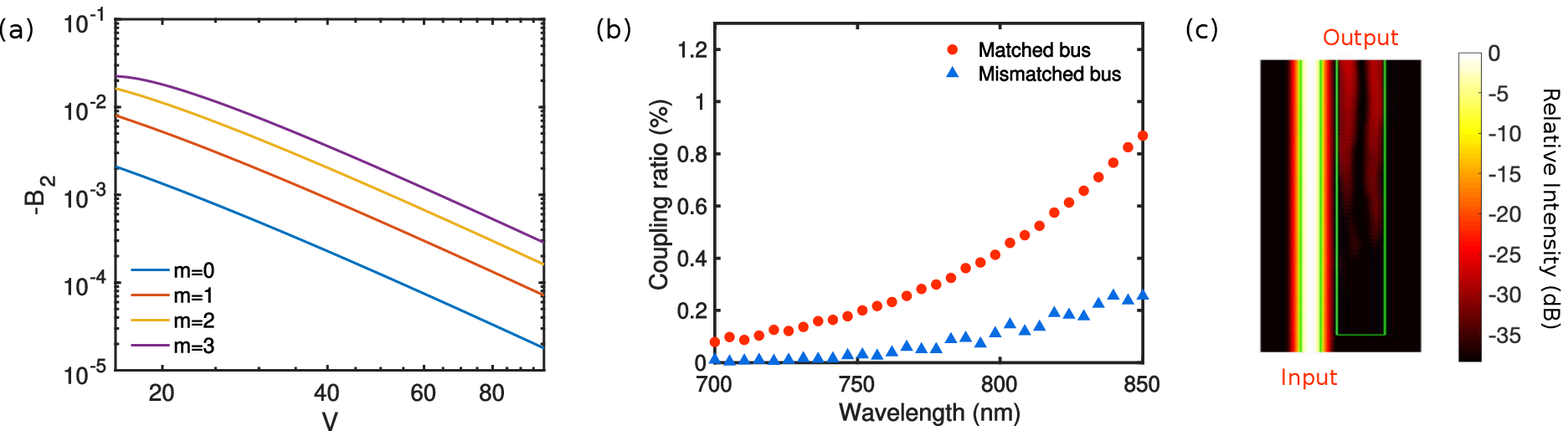}
	\caption{(a) Normalized GVD curve in log-log scale. (b) Simulated coupling bandwidth of different coupling schemes. (c) Field evolution of the coupling scheme, plotted as intensity in dB scale.}
	\label{figNormalizedParams}
\end{figure*}

We can also use the $B_2$ parameter to estimate the actual GVD scaling against the physical quantities such as waveguide height $h$ and index contrast $n_{co}^2-n_{cl}^2$. By plotting Fig. 1(a) of the main text on a log-log scale [Fig. \ref{figNormalizedParams}(a)],  we find that away from cutoff frequency, $B_2 \propto V^{-2.8}$, which, with Eqn. (2), leads to $B_2 \propto h^{-2.8}$ and $B_2 \propto -(n_{co}^2-n_{cl}^2)^{-1.4}$. Finally with Eqn. (4), we find that $\beta_{2,wg} \propto h^{-1.8}$ and $\beta_{2,wg} \propto (n_{co}^2-n_{cl}^2)^{0.1}$, which implies that for frequencies sufficiently away from cutoff value, the waveguide dispersion decreases strongly with an increase of waveguide size and is relatively insensitive to the change of refractive index contrast.

\section{Coupling scheme analysis}
We qualitatively show the bandwidth of our coupling scheme by simulating the coupling between two slab waveguides. The input waveguide has a width of 500 nm and the coupled waveguide has a width of 1200 nm such that the propagation constant of the TE$_0$ mode in the input waveguide matches that of the TE$_1$ mode in the coupled waveguide. We simulate a 16 $\mu$m long straight coupling region and find that the coupling strength can vary as much as 10 dB in a wavelength span of 150 nm. We also notice that the longer wavelength side has a larger coupling ratio. This is because at low coupling length regime, the coupling coefficient is more sensitive to mode overlap than wavevector mismatch and modes of longer wavelengths has more power in the evanescent field than that of the shorter wavelengths. However, a matched wavevector is also important as it guarantees an overall higher coupling efficiency and the excitation of the desired mode. To show this, we simulate the coupling ratio with a mismatched bus waveguide (375 nm). Even though the corresponding mode has more power in the evanescent field than the matched 500 nm bus, it has a significantly lower coupling ratio than the later. Finally, we note that this variation of coupling ratio is not unique to our mode converting coupling scheme. It exists in all broadband microresonantor combs.

\section{Device fabrication}
The devices are fabricated from a SiN thin film grown by low-pressure chemical vapor deposition (LPCVD). Electron beam lithography and reactive-ion etching are used to create the structures. We use inverse tapers at both ends of the bus waveguide to improve coupling efficiency and mode purity. The SiN waveguides are clad by high temperature oxide (HTO). Additionally, we implement platinum heaters above the cladding to assist Kerr comb generation (KCG) through thermal tuning \cite{Joshi_OL_2016}. 

\section{Experiment setup}
\begin{figure*}[htbp]
	\centering
	\includegraphics{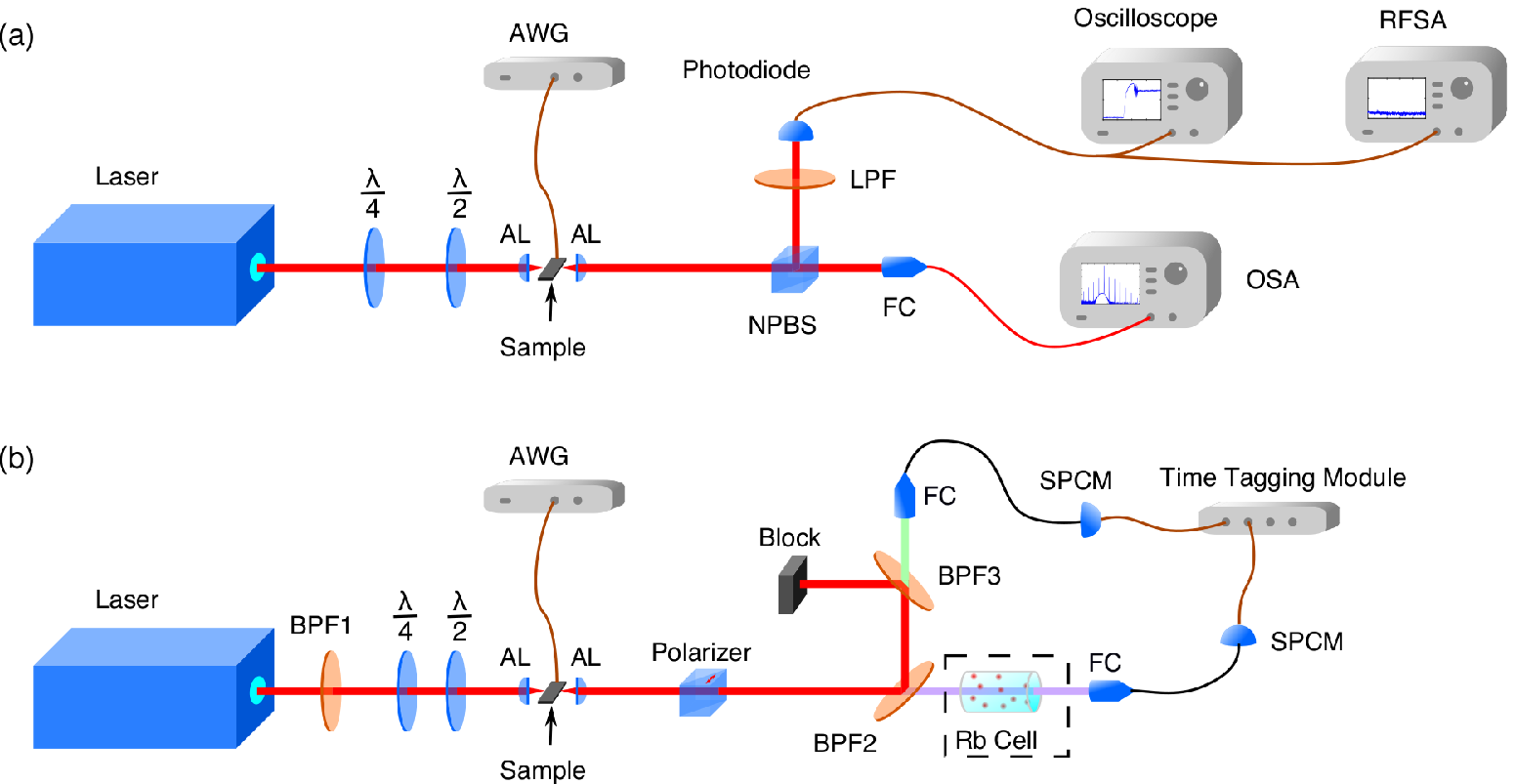}
	\caption{(a), (b), Experiment setup for KCG and PPG, respectively. AL, aspheric lens; AWG, arbitrary waveform generator; LPF, long pass filter with a cutoff at 800 nm. NPBS, nonpolarizing beam splitter; FC, fiber collimator; OSA, optical spectrum analyzer; RFSA, radio frequency spectrum analyzer; BPF1, BPF2, BPF3, band pass filter centered at 785 nm, 795 nm and 775 nm, respectively, with 3 nm bandwidth; SPCM, single photon counting module. }
	\label{figSetup}
\end{figure*}

\textbf{Experiment description for KCG.} Our laser source is an external cavity diode laser at 784 nm amplified by a tapered amplifier. The spatial mode is cleaned up by coupling into an optical fiber and subsequently coupled onto the chip through aspheric lenses. The input coupling loss is measured to be 4.4 dB with negligible propagation loss. We measure 220 mW of coupled power in the bus waveguide. An arbitrary waveform generator (AWG) is electrically connected to the on-chip heaters, which allows us to tune the resonance frequency through the thermal refractive effect. A higher heater power shifts the cavity resonance to the red side. We tune our laser frequency such that it is higher than the cavity resonance frequency at maximum heater power and lower than the cavity resonance frequency at room temperature. The comb is generated by scanning the heater voltage according to Fig. 2(a). The output light from the chip is collimated by an aspheric lens and subsequently split into two parts by a beam splitter. Half of the light is delivered into an optical spectrum analyzer (OSA) for spectral characterization and the other half is sent to a photodiode for power and radio frequency characterization.

For our comb interactions with a Rb cell, we use two bandpass filters (3-nm bandwidth) at 795 nm to pick up one comb line after the collection aspheric lens. We attenuate the power down to 10 $\mu$W to avoid saturation in absorption. We then place a 2-cm-long unshielded $^{85}$Rb cell after the filter and subsequently focus the transmitted light down onto a photodiode. The pump laser is wavelength modulated at 50 Hz while the photodiode output voltage is recorded on an oscilloscope.

\textbf{Experiment description for photon-pair generation (PPG).} We use the same amplified laser source as in KCG experiments. We use two bandpass filters (3-nm bandwidth) centered at 784.7 nm to reject the amplified spontaneous emission from the amplifier and spontaneous Raman scattering noise from the fiber before coupling the laser onto the chip. At the output we use a polarizer set for TE polarization to reject 50\% of the fluorescence noise. We then use similar bandpass filters to separate the desired signal (transmitted) and idler (reflected) photons. Further filtering is applied to provide enough pump rejection on both arms. Finally, the photons are coupled into optical fibers with black jackets. The filters create filtering losses and mode distortion, resulting in a total collection efficiency of 8.2\% for signal and 3.4\% for idler, which values are measured by driving the device above threshold. The photons are detected by SPCM's with 70\% efficiency and coincidences are measured by a time-tagging module (TTM), which is connected to the SPCM's.

For our photons interactions with a Rb cell, we place the cell in the signal path without altering the setup otherwise. The signal wavelength against pump wavelength is calibrated in the Kerr comb experiment. We lock our pump laser to a wavemeter with an accuracy of 0.01 pm. We use a pump power of 1.2 mW with a photon coincidence rate of 149 s$^{-1}$. We then tune the pump wavelength and measure coincidence rate after the cell for at least 2 minutes at each measurement point.

\section{Theory of continuous-wave laser pumped PPG}
A early theoretical and experimental investigation of PPG from a nonlinear cavity is performed by Ou and Lu \cite{Ou_PRL_1999}. Under low dispersion and low nonlinearity assumption in a $\chi^{(2)}$ cavity, they derived concise results for paired photon statistics. In this section, we use the same framework to study $\chi^{(3)}$ cavities and show that the same results apply, given that the parameters are properly redefined. In addition, we develop a full model incorporating dispersion, nonlinear phase and pump detuning and show that these are non-negligible factors in microresonantor based PPG.

Following \cite{Gardiner_PRA_1985}, the equation of motion for the signal and idler photons can be written as,
\begin{align}
\label{eqHeisenberg_s0} &\dot{\hat{a}}_s(t) = -i\omega_s \hat{a}_s(t) -\frac{i}{\hbar}[\hat{H}_{NL}(t), \hat{a}_s(t)] - \frac{\gamma_1 + \gamma_2}{2}\hat{a}_s(t) -\sqrt{\gamma_1}\hat{a}_{in}(t) -\sqrt{\gamma_2}\hat{b}_{in}(t)\\
\label{eqHeisenberg_i0}&\dot{\hat{a}}_i(t) = -\omega_i \hat{a}_i(t) -\frac{i}{\hbar}[\hat{H}_{NL}(t), \hat{a}_i(t)] - \frac{\gamma_1 + \gamma_2}{2}\hat{a}_i(t) -\sqrt{\gamma_1}\hat{a}_{in}(t) -\sqrt{\gamma_2}\hat{b}_{in}(t)
\end{align}
where $\hat{a}_s(t)$, $\hat{a}_i(t)$ are the photon annihilation operators for the signal and idler cavity modes, respectively, $\omega_s$, $\omega_i$ are the resonance frequencies for signal and idler, $\gamma_1$ and $\gamma_2$ are the bus-ring coupling rate and loss rate, $\hat{a}_{in}(t)$, $\hat{b}_{in}(t)$ are the corresponding Langevin operators for the two dissipation mechanisms. Under the classical pump assumption, we write the nonlinear Hamiltonian $\hat{H}_{NL}$ as,
\begin{align}
\hat{H}_{NL} = 2g |E_p|^2 (\hat{a}_s^\dagger \hat{a}_s + \hat{a}_i^\dagger \hat{a}_i) + g E_p^*E_p^* \hat{a}_s \hat{a}_i + H.c.
\end{align}
where $E_p$ is the pump field normalized so that $|E_p|^2$ is the classical circulating energy. $g = \hbar\gamma L/t_R^2$ where $\gamma$ is the nonlinear coefficient, $L$ is the round trip length and $t_R$ is the round trip time. Let $\Omega_p$ be the pump angular frequency, we make the following substitutions,
\begin{align}
&\Omega_s = \Omega_p-2\pi m/t_R\\
&\Omega_i = \Omega_p+2\pi m/t_R\\
&\hat{a}_s = e^{-i\Omega_s t}\int\hat{u}(\omega)e^{-i\omega t}d\omega\\
&\hat{a}_i = e^{-i\Omega_i t}\int\hat{v}(\omega)e^{-i\omega t}d\omega\\
&\hat{a}_{in} = e^{-i\Omega_s t}\int\hat{a}_{in}(\Omega_s+\omega)e^{-i\omega t}d\omega = e^{-i\Omega_i t}\int\hat{a}_{in}(\Omega_i+\omega)e^{-i\omega t}d\omega\\
&\hat{b}_{in} = e^{-i\Omega_s t}\int\hat{b}_{in}(\Omega_s+\omega)e^{-i\omega t}d\omega = e^{-i\Omega_i t}\int\hat{b}_{in}(\Omega_i+\omega)e^{-i\omega t}d\omega\\
&\epsilon = g|E_p|^2
\end{align}
where $m$ is the number of modes between the signal and the pump. Notice that $1/t_R$ is the free spectral range (FSR), hence $\Omega_s$ ($\Omega_i$) is the ``expected" angular frequency of the signal (idler). This allows us to incorporate pump detuning in our treatment as it is an important tuning parameter in experiments and applications. Equation (\ref{eqHeisenberg_s0})-(\ref{eqHeisenberg_i0}) can then be re-written in frequency domain as,
\begin{align} \label{eqHeisenberg_s1}
\begin{split}
\begin{pmatrix}
-i\omega & 0 \\
0 & -i\omega
\end{pmatrix}\begin{pmatrix}
\hat{u}(\omega) \\ \hat{v}^\dagger(-\omega)
\end{pmatrix} = &\begin{pmatrix}
i\Delta_s - \frac{\gamma_1 + \gamma_2}{2} +i2\epsilon & i\epsilon\\
-i\epsilon & -i\Delta_i - \frac{\gamma_1 + \gamma_2}{2} -i2\epsilon
\end{pmatrix}
\begin{pmatrix}
\hat{u}(\omega) \\ \hat{v}^\dagger(-\omega)
\end{pmatrix}\\ &- 
\sqrt{\gamma_1}\begin{pmatrix}
\hat{a}_{in}(\Omega_s+\omega) \\ \hat{a}_{in}^\dagger(\Omega_i-\omega)
\end{pmatrix} - 
\sqrt{\gamma_2}\begin{pmatrix}
\hat{b}_{in}(\Omega_s+\omega) \\ \hat{b}_{in}^\dagger(\Omega_i-\omega)
\end{pmatrix}
\end{split}
\end{align}
where $\Delta_s = \Omega_s-\omega_s$ and $\Delta_i = \Omega_i-\omega_i$ are the dispersion parameters as discussed later. Equation (\ref{eqHeisenberg_s1}) can be solved to find,
\begin{align}
\begin{split}
\begin{pmatrix}
\hat{u}(\omega) \\ \hat{v}^\dagger(-\omega)
\end{pmatrix} = &-\frac{1}{N}\begin{pmatrix}
-i(\omega-\Delta_i) + \frac{\gamma_1 + \gamma_2}{2} +i2\epsilon & i\epsilon\\
-i\epsilon & -i(\omega+\Delta_s) + \frac{\gamma_1 + \gamma_2}{2} -i2\epsilon 
\end{pmatrix}\\
&\times(\sqrt{\gamma_1}\begin{pmatrix}
\hat{a}_{in}(\Omega_s+\omega) \\ \hat{a}_{in}^\dagger(\Omega_i-\omega)
\end{pmatrix} + 
\sqrt{\gamma_2}\begin{pmatrix}
\hat{b}_{in}(\Omega_s+\omega) \\ \hat{b}_{in}^\dagger(\Omega_i-\omega)
\end{pmatrix})
\end{split}
\end{align}
where
\begin{align}
N = (\frac{\gamma_1+\gamma_2}{2}-i\frac{2\omega+\Delta_s-\Delta_i}{2})^2+(\frac{\Delta_s +\Delta_i}{2}+2\epsilon)^2-\epsilon^2
\end{align}
With the input-output relation, the output signal field is,
\begin{align}
\hat{a}_{out}(\Omega_s+\omega) = \sqrt{\gamma_1}\hat{u}(\omega) + \hat{a}_{in}(\Omega_s + \omega)
\end{align}
We get,
\begin{align}\label{eqSignalOut}
\begin{split}
\hat{a}_{out}(\Omega_s+\omega) = &G_1(\omega)\hat{a}_{in}(\Omega_s+\omega) + g_1(\omega)\hat{a}_{in}^\dagger(\Omega_i-\omega)\\
&+G_2(\omega)\hat{b}_{in}(\Omega_s+\omega) + g_2(\omega)\hat{b}_{in}^\dagger(\Omega_i-\omega)
\end{split}
\end{align}
with,
\begin{align}
&G_1(\omega) =  -\frac{[\gamma_1+\gamma_2-i(2\omega+\Delta_s-\Delta_i)][\gamma_1-\gamma_2+i(2\omega+\Delta_s-\Delta_i)]}{[(\gamma_1+\gamma_2)-i(2\omega+\Delta_s-\Delta_i)]^2+[(\Delta_s +\Delta_i)+4\epsilon]^2-4\epsilon^2} \notag\\
&\qquad\qquad+ \frac{-i2\gamma_1(\Delta_s+\Delta_i+4\epsilon)+[(\Delta_s +\Delta_i)+4\epsilon]^2-4\epsilon^2}{[(\gamma_1+\gamma_2)-i(2\omega+\Delta_s-\Delta_i)]^2+[(\Delta_s +\Delta_i)+4\epsilon]^2-4\epsilon^2}\\
&g_1(\omega) = -\frac{i4\epsilon \gamma_1}{[(\gamma_1+\gamma_2)-i(2\omega+\Delta_s-\Delta_i)]^2+[(\Delta_s +\Delta_i)+4\epsilon]^2-4\epsilon^2}\\
&G_2(\omega) = -\frac{2\sqrt{\gamma_1\gamma_2}[(\gamma_1+\gamma_2-i(2\omega+\Delta_s-\Delta_i)+i(\Delta_s+\Delta_i)+i4\epsilon]}{[(\gamma_1+\gamma_2)-i(2\omega+\Delta_s-\Delta_i)]^2+[(\Delta_s +\Delta_i)+4\epsilon]^2-4\epsilon^2}\\
&g_2(\omega) = -\frac{i4\epsilon \sqrt{\gamma_1\gamma_2}}{[(\gamma_1+\gamma_2)-i(2\omega+\Delta_s-\Delta_i)]^2+[(\Delta_s +\Delta_i)+4\epsilon]^2-4\epsilon^2}
\end{align}
Similarly,
\begin{align}\label{eqIdlerOut}
\begin{split}
\hat{a}_{out}(\Omega_i-\omega) = &G_1(-\omega)\hat{a}_{in}(\Omega_i-\omega) + g_1(-\omega)\hat{a}_{in}^\dagger(\Omega_s+\omega)\\
&+G_2(-\omega)\hat{b}_{in}(\Omega_i-\omega) + g_2(-\omega)\hat{b}_{in}^\dagger(\Omega_s+\omega)
\end{split}
\end{align}
At the limit of low dispersion, low nonlinearity and low pump detuning, that is $\Delta_s, \Delta_i, \epsilon \ll \gamma_1, \gamma_2$,  Eqn. (\ref{eqSignalOut}) and (\ref{eqIdlerOut}) are reduced to Eqn. (1) of \cite{Ou_PRL_1999} with the following two differences. First, there is a $\pi$ phase difference between our result and that in \cite{Ou_PRL_1999}. This is because in a Fabry-P\'erot cavity considered in \cite{Ou_PRL_1999} the input and output fields have a $\pi$ phase difference due to mirror reflection while in our ring resonator this phase is absent for the fields in the bus waveguide. Second, \cite{Ou_PRL_1999} focused on PPG within the same resonance as the pump while we show that the treatment can be generalized to treat signal and idler in different resonances, as is the case in our experiment. As is shown in \cite{Ou_PRL_1999}, we write down the full width at half maximum (FWHM) bandwidth of the photon spectrum, without dispersion, to be,
\begin{align}
\Delta\omega = 0.64(\gamma_1+\gamma_2)
\end{align}
and the FWHM correlation time to be,
\begin{align}
T_c = \frac{1.39}{\gamma_1+\gamma_2}
\end{align}
To understand the photon spectrum in this experiment, we need to return to our full model, which yields a spectrum,
\begin{align}\label{eqSpectrum2}
S(\Omega_s+\omega) &= |g_1(\omega)|^2+|g_2(\omega)|^2 \notag\\
&=\frac{16\epsilon^2\gamma_1(\gamma_1+\gamma_2)}{[(2\omega+\Delta_s-\Delta_i)^2+(\gamma_1+\gamma_2)^2-(\Delta_s+\Delta_i+4\epsilon)^2+4\epsilon^2]^2+4(\gamma_1+\gamma_2)^2[(\Delta_s+\Delta_i+4\epsilon)^2-4\epsilon^2]}
\end{align}

\begin{figure}
	\centering
	\includegraphics[width=\linewidth]{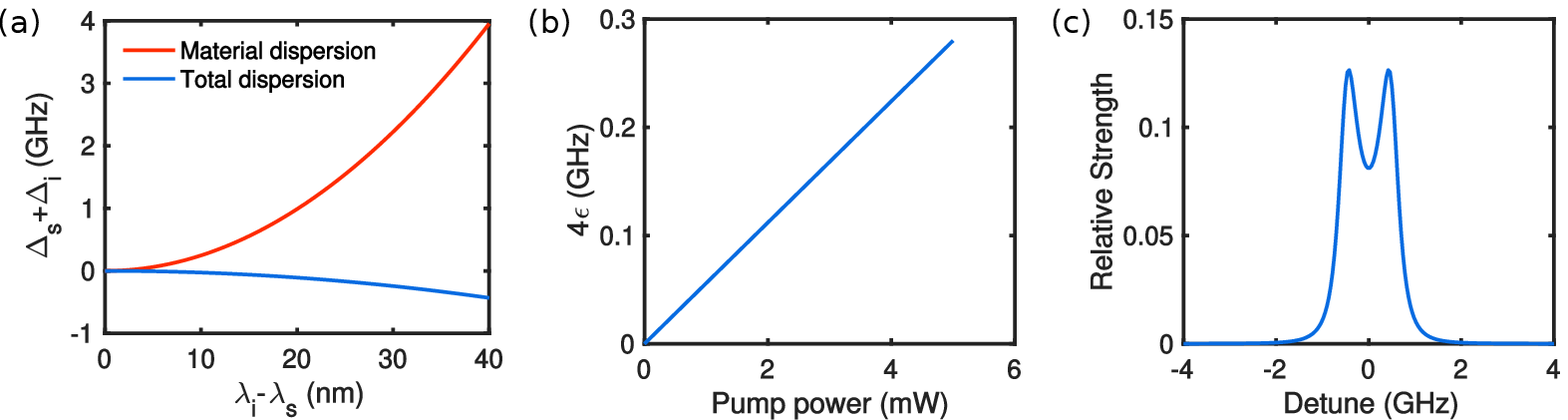}
	\caption{(a) Scaling of the dispersion parameter in Eqn. (\ref{eqSpectrum2}). (b) Scaling of the nonlinear phase parameter in Eqn. (\ref{eqSpectrum2}). (c) Infered photon spectrum generated in the experiment.}
	\label{figPhotonTheory}
\end{figure}
We link the parameters in Eqn. (\ref{eqSpectrum}) to experimentally accessible values. Up to second order dispersion, we have,
\begin{align}
&\gamma_2 = \omega_p/Q_i\\
&\gamma_1+\gamma_2 = \omega_p/Q_l\\
&\Delta_s-\Delta_i \approx 0\\
&\Delta_s+\Delta_i = 2\Delta_p + \frac{\beta_2}{\beta_1}(2\pi mf_{\text{FSR}})^2 + \Delta_{\text{MI}}\\
&\epsilon \approx \gamma PLf_{\text{FSR}}\frac{4\gamma_1 f_{\text{FSR}}}{\Delta_p^2+(\gamma_1+\gamma_2)^2}
\end{align}
where $Q_i$ and $Q_l$ are the intrinsic and loaded $Q$, respectively, $\beta_1$ and $\beta_2$ are first and second order dispersion coefficients at pump wavelength, $\omega_p$ is the pump resonance angular frequency, $\Delta_p = \Omega_p-\omega_p$ is pump detuning, $P$ is input pump power in the bus waveguide, $f_{\text{FSR}} = 1/t_R$ is the FSR, $\Delta_{\text{MI}}$ is the resonance shift caused by mode interactions and $L$ is the round trip length. From Eqn. (\ref{eqSpectrum}) we see that the detuning and dispersion effect is contained in the parameter $\Delta_s+\Delta_i$ and the nonlinear cross phase modulation is contained in the parameter $4\epsilon$. The low dispersion and low nonlinearity limit requires that $\Delta_s+\Delta_i, 4\epsilon \ll \gamma_1+\gamma_2$. However, by plotting these parameters for our device (Fig. \ref{figPhotonTheory}), we notice that their effect is non-negligible for photons far from the pump. Additionally, if the pump is blue detuned ($\Delta_p > 0$), which is the soft thermal locking regime, it can counteract with anomalous GVD to reduce the overall dispersion effect. Intuitively, this is because that energy conservation requires signal and idler spectra be symmetric with respect to the pump but anomalous GVD shifts both signal and idler resonance to the blue side so that they are asymmetric with respect to the pump resonance. However, they can be symmetric with respect to a blue detuned pump, in which case the dispersion effect is minimized. As mentioned in the main text, Eqn. (\ref{eqSpectrum2}) is used to fit the Rb absorption spectrum and the best fit photon spectrum is plotted in Fig. \ref{figPhotonTheory}(c). The double-peak feature is caused by a cavity mode interaction around 795 nm, which pushes the signal resonance away from the symmetry point of the idler resonance. As a result, the photon spectrum shows two peaks corresponding to the two shifted resonances.

\begin{figure}[htbp]
	\centering
	\includegraphics{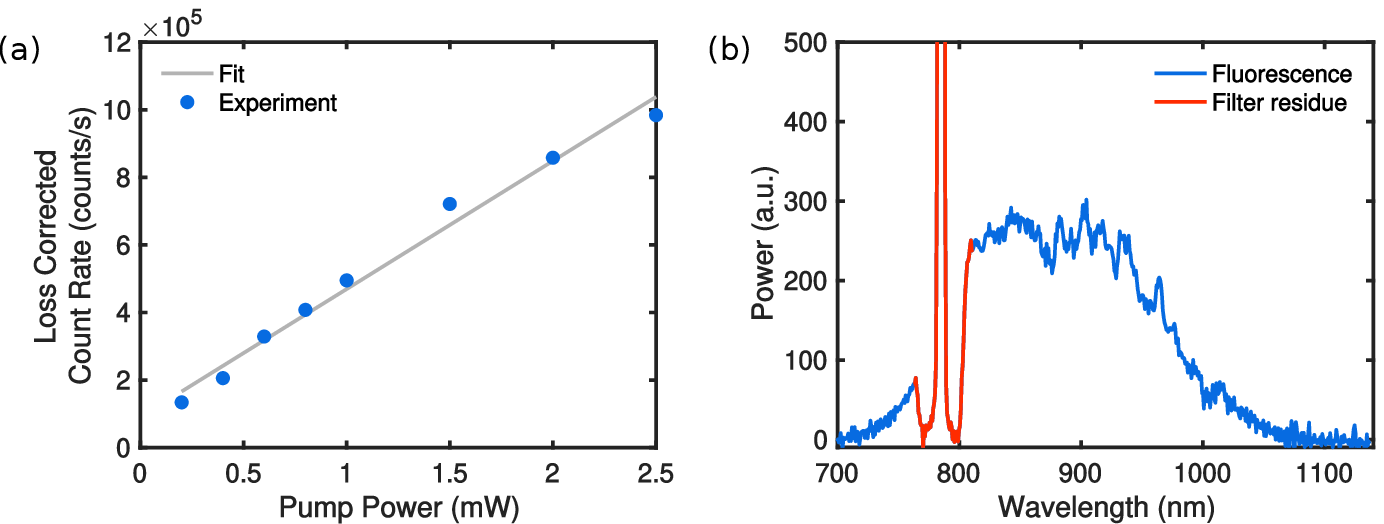}
	\caption{Fluorescence characterization. (a) Loss corrected noise photon count rate in experiment. (b) Spectrum of fluorescence.}
	\label{figPhotonNoise}
\end{figure}

\section{Fluorescence from silicon nitride}
We first characterize the noise photon behavior by measuring the noise photon count rate in the signal arm. The pump is set off resonance and its power is varied. We plot the noise photon count rate against pump power (Fig. \ref{figPhotonNoise}a) and find that the scaling is linear, as is expected for fluorescence. We also note that when pump is tuned into resonance, extra Purcell-effect-enhanced fluorescence is produced within the resonance bandwidth. 

Next we measure the spectrum of the noise photons with a low-light-level spectrometer. We set the off resonance pump to the TE mode of the bus waveguide and collect fluorescence from the TM mode with a polarizer. An additional notch filter is used to further reject the pump. We collect light with a multimode fiber to minimize the wavelength dependent coupling variation. The spectral measurement confirms the existence of broad band fluorescence on both sides of the pump. As expected, the lower photon energy side produces stronger fluorescence than the high photon energy side. This fluorescence can be reduced by reducing the silicon concentration in the waveguide material.


\begin{thebibliography}{55}%
	\makeatletter
	\providecommand \@ifxundefined [1]{%
		\@ifx{#1\undefined}
	}%
	\providecommand \@ifnum [1]{%
		\ifnum #1\expandafter \@firstoftwo
		\else \expandafter \@secondoftwo
		\fi
	}%
	\providecommand \@ifx [1]{%
		\ifx #1\expandafter \@firstoftwo
		\else \expandafter \@secondoftwo
		\fi
	}%
	\providecommand \natexlab [1]{#1}%
	\providecommand \enquote  [1]{``#1''}%
	\providecommand \bibnamefont  [1]{#1}%
	\providecommand \bibfnamefont [1]{#1}%
	\providecommand \citenamefont [1]{#1}%
	\providecommand \href@noop [0]{\@secondoftwo}%
	\providecommand \href [0]{\begingroup \@sanitize@url \@href}%
	\providecommand \@href[1]{\@@startlink{#1}\@@href}%
	\providecommand \@@href[1]{\endgroup#1\@@endlink}%
	\providecommand \@sanitize@url [0]{\catcode `\\12\catcode `\$12\catcode
		`\&12\catcode `\#12\catcode `\^12\catcode `\_12\catcode `\%12\relax}%
	\providecommand \@@startlink[1]{}%
	\providecommand \@@endlink[0]{}%
	\providecommand \url  [0]{\begingroup\@sanitize@url \@url }%
	\providecommand \@url [1]{\endgroup\@href {#1}{\urlprefix }}%
	\providecommand \urlprefix  [0]{URL }%
	\providecommand \Eprint [0]{\href }%
	\providecommand \doibase [0]{https://doi.org/}%
	\providecommand \selectlanguage [0]{\@gobble}%
	\providecommand \bibinfo  [0]{\@secondoftwo}%
	\providecommand \bibfield  [0]{\@secondoftwo}%
	\providecommand \translation [1]{[#1]}%
	\providecommand \BibitemOpen [0]{}%
	\providecommand \bibitemStop [0]{}%
	\providecommand \bibitemNoStop [0]{.\EOS\space}%
	\providecommand \EOS [0]{\spacefactor3000\relax}%
	\providecommand \BibitemShut  [1]{\csname bibitem#1\endcsname}%
	\let\auto@bib@innerbib\@empty
	\bibitem [{\citenamefont {Turner}\ \emph {et~al.}(2006)\citenamefont {Turner},
		\citenamefont {Manolatou}, \citenamefont {Schmidt}, \citenamefont {Lipson},
		\citenamefont {Foster}, \citenamefont {Sharping},\ and\ \citenamefont
		{Gaeta}}]{Turner_OE_2006}%
	\BibitemOpen
	\bibfield  {author} {\bibinfo {author} {\bibfnamefont {A.~C.}\ \bibnamefont
			{Turner}}, \bibinfo {author} {\bibfnamefont {C.}~\bibnamefont {Manolatou}},
		\bibinfo {author} {\bibfnamefont {B.~S.}\ \bibnamefont {Schmidt}}, \bibinfo
		{author} {\bibfnamefont {M.}~\bibnamefont {Lipson}}, \bibinfo {author}
		{\bibfnamefont {M.~A.}\ \bibnamefont {Foster}}, \bibinfo {author}
		{\bibfnamefont {J.~E.}\ \bibnamefont {Sharping}}, and\ \bibinfo {author}
		{\bibfnamefont {A.~L.}\ \bibnamefont {Gaeta}},\ }\bibfield  {title} {\bibinfo
		{title} {Tailored anomalous group-velocity dispersion in silicon channel
			waveguides},\ }\href {https://doi.org/10.1364/OE.14.004357} {\bibfield
		{journal} {\bibinfo  {journal} {Opt. Express}\ }\textbf {\bibinfo {volume}
			{14}},\ \bibinfo {pages} {4357} (\bibinfo {year} {2006})}\BibitemShut
	{NoStop}%
	\bibitem [{\citenamefont {Kippenberg}\ \emph {et~al.}(2018)\citenamefont
		{Kippenberg}, \citenamefont {Gaeta}, \citenamefont {Lipson},\ and\
		\citenamefont {Gorodetsky}}]{Kippenberg_Science_2018}%
	\BibitemOpen
	\bibfield  {author} {\bibinfo {author} {\bibfnamefont {T.~J.}\ \bibnamefont
			{Kippenberg}}, \bibinfo {author} {\bibfnamefont {A.~L.}\ \bibnamefont
			{Gaeta}}, \bibinfo {author} {\bibfnamefont {M.}~\bibnamefont {Lipson}}, and\
		\bibinfo {author} {\bibfnamefont {M.~L.}\ \bibnamefont {Gorodetsky}},\
	}\bibfield  {title} {\bibinfo {title} {Dissipative kerr solitons in optical
			microresonators},\ }\href {https://doi.org/10.1126/science.aan8083}
	{\bibfield  {journal} {\bibinfo  {journal} {Science}\ }\textbf {\bibinfo
			{volume} {361}},\ \bibinfo {pages} {eaan8083} (\bibinfo {year}
		{2018})}\BibitemShut {NoStop}%
	\bibitem [{\citenamefont {Gaeta}\ \emph {et~al.}(2019)\citenamefont {Gaeta},
		\citenamefont {Lipson},\ and\ \citenamefont
		{Kippenberg}}]{Gaeta_NatPhot_2019}%
	\BibitemOpen
	\bibfield  {author} {\bibinfo {author} {\bibfnamefont {A.~L.}\ \bibnamefont
			{Gaeta}}, \bibinfo {author} {\bibfnamefont {M.}~\bibnamefont {Lipson}}, and\
		\bibinfo {author} {\bibfnamefont {T.~J.}\ \bibnamefont {Kippenberg}},\
	}\bibfield  {title} {\bibinfo {title} {Photonic-chip-based frequency combs},\
	}\href {https://doi.org/10.1038/s41566-019-0358-x} {\bibfield  {journal}
		{\bibinfo  {journal} {Nat. Photonics}\ }\textbf {\bibinfo {volume} {13}},\
		\bibinfo {pages} {158} (\bibinfo {year} {2019})}\BibitemShut {NoStop}%
	\bibitem [{\citenamefont {Kues}\ \emph {et~al.}(2019)\citenamefont {Kues},
		\citenamefont {Reimer}, \citenamefont {Lukens}, \citenamefont {Munro},
		\citenamefont {Weiner}, \citenamefont {Moss},\ and\ \citenamefont
		{Morandotti}}]{Kues_NatPhot_2019}%
	\BibitemOpen
	\bibfield  {author} {\bibinfo {author} {\bibfnamefont {M.}~\bibnamefont
			{Kues}}, \bibinfo {author} {\bibfnamefont {C.}~\bibnamefont {Reimer}},
		\bibinfo {author} {\bibfnamefont {J.~M.}\ \bibnamefont {Lukens}}, \bibinfo
		{author} {\bibfnamefont {W.~J.}\ \bibnamefont {Munro}}, \bibinfo {author}
		{\bibfnamefont {A.~M.}\ \bibnamefont {Weiner}}, \bibinfo {author}
		{\bibfnamefont {D.~J.}\ \bibnamefont {Moss}}, and\ \bibinfo {author}
		{\bibfnamefont {R.}~\bibnamefont {Morandotti}},\ }\bibfield  {title}
	{\bibinfo {title} {Quantum optical microcombs},\ }\href
	{https://doi.org/10.1038/s41566-019-0363-0} {\bibfield  {journal} {\bibinfo
			{journal} {Nat. Photonics}\ }\textbf {\bibinfo {volume} {13}},\ \bibinfo
		{pages} {170} (\bibinfo {year} {2019})}\BibitemShut {NoStop}%
	\bibitem [{\citenamefont {Boyd}(2008)}]{Boyd_NO}%
	\BibitemOpen
	\bibfield  {author} {\bibinfo {author} {\bibfnamefont {R.~W.}\ \bibnamefont
			{Boyd}},\ }\href@noop {} {\emph {\bibinfo {title} {Nonlinear Optics}}},\
	\bibinfo {edition} {3rd}\ ed.\ (\bibinfo  {publisher} {Academic Press Inc.},\
	\bibinfo {address} {Orlando, FL, USA},\ \bibinfo {year} {2008})\BibitemShut
	{NoStop}%
	\bibitem [{\citenamefont {Pollock}\ and\ \citenamefont
		{Lipson}(2003)}]{Pollock_IP}%
	\BibitemOpen
	\bibfield  {author} {\bibinfo {author} {\bibfnamefont {C.~R.}\ \bibnamefont
			{Pollock}}and\ \bibinfo {author} {\bibfnamefont {M.}~\bibnamefont {Lipson}},\
	}\href@noop {} {\emph {\bibinfo {title} {Integrated Photonics}}}\ (\bibinfo
	{publisher} {Kluwer Academic Publishers},\ \bibinfo {address} {Norwell, MA,
		USA},\ \bibinfo {year} {2003})\BibitemShut {NoStop}%
	\bibitem [{\citenamefont {Bennett}\ and\ \citenamefont
		{Chen}(1980)}]{Bennett_AO_1980}%
	\BibitemOpen
	\bibfield  {author} {\bibinfo {author} {\bibfnamefont {G.~A.}\ \bibnamefont
			{Bennett}}and\ \bibinfo {author} {\bibfnamefont {C.-L.}\ \bibnamefont
			{Chen}},\ }\bibfield  {title} {\bibinfo {title} {Wavelength dispersion of
			optical waveguides},\ }\href {https://doi.org/10.1364/AO.19.001990}
	{\bibfield  {journal} {\bibinfo  {journal} {Appl. Opt.}\ }\textbf {\bibinfo
			{volume} {19}},\ \bibinfo {pages} {1990} (\bibinfo {year}
		{1980})}\BibitemShut {NoStop}%
	\bibitem [{\citenamefont {Savchenkov}\ \emph {et~al.}(2011)\citenamefont
		{Savchenkov}, \citenamefont {Matsko}, \citenamefont {Liang}, \citenamefont
		{Ilchenko}, \citenamefont {Seidel},\ and\ \citenamefont
		{Maleki}}]{Savchenkov_NatPhot_2011}%
	\BibitemOpen
	\bibfield  {author} {\bibinfo {author} {\bibfnamefont {A.~A.}\ \bibnamefont
			{Savchenkov}}, \bibinfo {author} {\bibfnamefont {A.~B.}\ \bibnamefont
			{Matsko}}, \bibinfo {author} {\bibfnamefont {W.}~\bibnamefont {Liang}},
		\bibinfo {author} {\bibfnamefont {V.~S.}\ \bibnamefont {Ilchenko}}, \bibinfo
		{author} {\bibfnamefont {D.}~\bibnamefont {Seidel}}, and\ \bibinfo {author}
		{\bibfnamefont {L.}~\bibnamefont {Maleki}},\ }\bibfield  {title} {\bibinfo
		{title} {Kerr combs with selectable central frequency},\ }\href
	{https://doi.org/10.1038/nphoton.2011.50} {\bibfield  {journal} {\bibinfo
			{journal} {Nat. Photonics}\ }\textbf {\bibinfo {volume} {5}},\ \bibinfo
		{pages} {293} (\bibinfo {year} {2011})}\BibitemShut {NoStop}%
	\bibitem [{\citenamefont {Luke}\ \emph {et~al.}(2015)\citenamefont {Luke},
		\citenamefont {Okawachi}, \citenamefont {Lamont}, \citenamefont {Gaeta},\
		and\ \citenamefont {Lipson}}]{Luke_OL_2015}%
	\BibitemOpen
	\bibfield  {author} {\bibinfo {author} {\bibfnamefont {K.}~\bibnamefont
			{Luke}}, \bibinfo {author} {\bibfnamefont {Y.}~\bibnamefont {Okawachi}},
		\bibinfo {author} {\bibfnamefont {M.~R.~E.}\ \bibnamefont {Lamont}}, \bibinfo
		{author} {\bibfnamefont {A.~L.}\ \bibnamefont {Gaeta}}, and\ \bibinfo
		{author} {\bibfnamefont {M.}~\bibnamefont {Lipson}},\ }\bibfield  {title}
	{\bibinfo {title} {Broadband mid-infrared frequency comb generation in a
			si3n4 microresonator},\ }\href {https://doi.org/10.1364/OL.40.004823}
	{\bibfield  {journal} {\bibinfo  {journal} {Opt. Lett.}\ }\textbf {\bibinfo
			{volume} {40}},\ \bibinfo {pages} {4823} (\bibinfo {year}
		{2015})}\BibitemShut {NoStop}%
	\bibitem [{\citenamefont {Luo}\ \emph {et~al.}(2014)\citenamefont {Luo},
		\citenamefont {Ophir}, \citenamefont {Chen}, \citenamefont {Gabrielli},
		\citenamefont {Poitras}, \citenamefont {Bergmen},\ and\ \citenamefont
		{Lipson}}]{Luo_NatComm_2014}%
	\BibitemOpen
	\bibfield  {author} {\bibinfo {author} {\bibfnamefont {L.-W.}\ \bibnamefont
			{Luo}}, \bibinfo {author} {\bibfnamefont {N.}~\bibnamefont {Ophir}}, \bibinfo
		{author} {\bibfnamefont {C.~P.}\ \bibnamefont {Chen}}, \bibinfo {author}
		{\bibfnamefont {L.~H.}\ \bibnamefont {Gabrielli}}, \bibinfo {author}
		{\bibfnamefont {C.~B.}\ \bibnamefont {Poitras}}, \bibinfo {author}
		{\bibfnamefont {K.}~\bibnamefont {Bergmen}}, and\ \bibinfo {author}
		{\bibfnamefont {M.}~\bibnamefont {Lipson}},\ }\bibfield  {title} {\bibinfo
		{title} {{WDM}-compatible mode-division multiplexing on a silicon chip},\
	}\href {https://doi.org/10.1038/ncomms4069} {\bibfield  {journal} {\bibinfo
			{journal} {Nat. Commun.}\ }\textbf {\bibinfo {volume} {5}},\ \bibinfo {pages}
		{3069} (\bibinfo {year} {2014})}\BibitemShut {NoStop}%
	\bibitem [{\citenamefont {Herr}\ \emph {et~al.}(2014)\citenamefont {Herr},
		\citenamefont {Brasch}, \citenamefont {Jost}, \citenamefont {Wang},
		\citenamefont {Kondratiev}, \citenamefont {Gorodetsky},\ and\ \citenamefont
		{Kippenberg}}]{Herr_NatPhot_2014}%
	\BibitemOpen
	\bibfield  {author} {\bibinfo {author} {\bibfnamefont {T.}~\bibnamefont
			{Herr}}, \bibinfo {author} {\bibfnamefont {V.}~\bibnamefont {Brasch}},
		\bibinfo {author} {\bibfnamefont {J.~D.}\ \bibnamefont {Jost}}, \bibinfo
		{author} {\bibfnamefont {C.~Y.}\ \bibnamefont {Wang}}, \bibinfo {author}
		{\bibfnamefont {N.~M.}\ \bibnamefont {Kondratiev}}, \bibinfo {author}
		{\bibfnamefont {M.~L.}\ \bibnamefont {Gorodetsky}}, and\ \bibinfo {author}
		{\bibfnamefont {T.~J.}\ \bibnamefont {Kippenberg}},\ }\bibfield  {title}
	{\bibinfo {title} {Temporal solitons in optical microresonators},\ }\href
	{https://doi.org/10.1038/nphoton.2013.343} {\bibfield  {journal} {\bibinfo
			{journal} {Nat. Photonics}\ }\textbf {\bibinfo {volume} {8}},\ \bibinfo
		{pages} {145} (\bibinfo {year} {2014})}\BibitemShut {NoStop}%
	\bibitem [{\citenamefont {Stern}\ \emph {et~al.}(2018)\citenamefont {Stern},
		\citenamefont {Ji}, \citenamefont {Okawachi}, \citenamefont {Gaeta},\ and\
		\citenamefont {Lipson}}]{Stern_Nature_2018}%
	\BibitemOpen
	\bibfield  {author} {\bibinfo {author} {\bibfnamefont {B.}~\bibnamefont
			{Stern}}, \bibinfo {author} {\bibfnamefont {X.}~\bibnamefont {Ji}}, \bibinfo
		{author} {\bibfnamefont {Y.}~\bibnamefont {Okawachi}}, \bibinfo {author}
		{\bibfnamefont {A.~L.}\ \bibnamefont {Gaeta}}, and\ \bibinfo {author}
		{\bibfnamefont {M.}~\bibnamefont {Lipson}},\ }\bibfield  {title} {\bibinfo
		{title} {Battery-operated integrated frequency comb generator},\ }\href
	{https://doi.org/10.1038/s41586-018-0598-9} {\bibfield  {journal} {\bibinfo
			{journal} {Nature}\ }\textbf {\bibinfo {volume} {562}},\ \bibinfo {pages}
		{401} (\bibinfo {year} {2018})}\BibitemShut {NoStop}%
	\bibitem [{\citenamefont {Suh}\ \emph {et~al.}(2016)\citenamefont {Suh},
		\citenamefont {Yang}, \citenamefont {Yang}, \citenamefont {Yi},\ and\
		\citenamefont {Vahala}}]{Suh_Science_2016}%
	\BibitemOpen
	\bibfield  {author} {\bibinfo {author} {\bibfnamefont {M.-G.}\ \bibnamefont
			{Suh}}, \bibinfo {author} {\bibfnamefont {Q.-F.}\ \bibnamefont {Yang}},
		\bibinfo {author} {\bibfnamefont {K.~Y.}\ \bibnamefont {Yang}}, \bibinfo
		{author} {\bibfnamefont {X.}~\bibnamefont {Yi}}, and\ \bibinfo {author}
		{\bibfnamefont {K.~J.}\ \bibnamefont {Vahala}},\ }\bibfield  {title}
	{\bibinfo {title} {Microresonator soliton dual-comb spectroscopy},\ }\href
	{https://doi.org/10.1126/science.aah6516} {\bibfield  {journal} {\bibinfo
			{journal} {Science}\ }\textbf {\bibinfo {volume} {354}},\ \bibinfo {pages}
		{600} (\bibinfo {year} {2016})}\BibitemShut {NoStop}%
	\bibitem [{\citenamefont {Pavlov}\ \emph {et~al.}(2017)\citenamefont {Pavlov},
		\citenamefont {Lihachev}, \citenamefont {Koptyaev}, \citenamefont {Lucas},
		\citenamefont {Karpov}, \citenamefont {Kondratiev}, \citenamefont {Bilenko},
		\citenamefont {Kippenberg},\ and\ \citenamefont
		{Gorodetsky}}]{Pavlov_OL_2017}%
	\BibitemOpen
	\bibfield  {author} {\bibinfo {author} {\bibfnamefont {N.~G.}\ \bibnamefont
			{Pavlov}}, \bibinfo {author} {\bibfnamefont {G.}~\bibnamefont {Lihachev}},
		\bibinfo {author} {\bibfnamefont {S.}~\bibnamefont {Koptyaev}}, \bibinfo
		{author} {\bibfnamefont {E.}~\bibnamefont {Lucas}}, \bibinfo {author}
		{\bibfnamefont {M.}~\bibnamefont {Karpov}}, \bibinfo {author} {\bibfnamefont
			{N.~M.}\ \bibnamefont {Kondratiev}}, \bibinfo {author} {\bibfnamefont
			{I.~A.}\ \bibnamefont {Bilenko}}, \bibinfo {author} {\bibfnamefont {T.~J.}\
			\bibnamefont {Kippenberg}}, and\ \bibinfo {author} {\bibfnamefont {M.~L.}\
			\bibnamefont {Gorodetsky}},\ }\bibfield  {title} {\bibinfo {title} {Soliton
			dual frequency combs in crystalline microresonators},\ }\href
	{https://doi.org/10.1364/OL.42.000514} {\bibfield  {journal} {\bibinfo
			{journal} {Opt. Lett.}\ }\textbf {\bibinfo {volume} {42}},\ \bibinfo {pages}
		{514} (\bibinfo {year} {2017})}\BibitemShut {NoStop}%
	\bibitem [{\citenamefont {Yu}\ \emph {et~al.}(2018)\citenamefont {Yu},
		\citenamefont {Okawachi}, \citenamefont {Griffith}, \citenamefont {Picqué},
		\citenamefont {Lipson},\ and\ \citenamefont {Gaeta}}]{Yu_NatComm_2018}%
	\BibitemOpen
	\bibfield  {author} {\bibinfo {author} {\bibfnamefont {M.}~\bibnamefont
			{Yu}}, \bibinfo {author} {\bibfnamefont {Y.}~\bibnamefont {Okawachi}},
		\bibinfo {author} {\bibfnamefont {A.~G.}\ \bibnamefont {Griffith}}, \bibinfo
		{author} {\bibfnamefont {N.}~\bibnamefont {Picqué}}, \bibinfo {author}
		{\bibfnamefont {M.}~\bibnamefont {Lipson}}, and\ \bibinfo {author}
		{\bibfnamefont {A.~L.}\ \bibnamefont {Gaeta}},\ }\bibfield  {title} {\bibinfo
		{title} {Silicon-chip-based mid-infrared dual-comb spectroscopy},\ }\href
	{https://doi.org/10.1038/s41467-018-04350-1} {\bibfield  {journal} {\bibinfo
			{journal} {Nat. Commun.}\ }\textbf {\bibinfo {volume} {9}},\ \bibinfo {pages}
		{1869} (\bibinfo {year} {2018})}\BibitemShut {NoStop}%
	\bibitem [{\citenamefont {Trocha}\ \emph {et~al.}(2018)\citenamefont {Trocha},
		\citenamefont {Karpov}, \citenamefont {Ganin}, \citenamefont {Pfeiffer},
		\citenamefont {Kordts}, \citenamefont {Wolf}, \citenamefont {Krockenberger},
		\citenamefont {Marin-Palomo}, \citenamefont {Weimann}, \citenamefont
		{Randel}, \citenamefont {Freude}, \citenamefont {Kippenberg},\ and\
		\citenamefont {Koos}}]{Trocha_Science_2018}%
	\BibitemOpen
	\bibfield  {author} {\bibinfo {author} {\bibfnamefont {P.}~\bibnamefont
			{Trocha}}, \bibinfo {author} {\bibfnamefont {M.}~\bibnamefont {Karpov}},
		\bibinfo {author} {\bibfnamefont {D.}~\bibnamefont {Ganin}}, \bibinfo
		{author} {\bibfnamefont {M.~H.~P.}\ \bibnamefont {Pfeiffer}}, \bibinfo
		{author} {\bibfnamefont {A.}~\bibnamefont {Kordts}}, \bibinfo {author}
		{\bibfnamefont {S.}~\bibnamefont {Wolf}}, \bibinfo {author} {\bibfnamefont
			{J.}~\bibnamefont {Krockenberger}}, \bibinfo {author} {\bibfnamefont
			{P.}~\bibnamefont {Marin-Palomo}}, \bibinfo {author} {\bibfnamefont
			{C.}~\bibnamefont {Weimann}}, \bibinfo {author} {\bibfnamefont
			{S.}~\bibnamefont {Randel}}, \bibinfo {author} {\bibfnamefont
			{W.}~\bibnamefont {Freude}}, \bibinfo {author} {\bibfnamefont {T.~J.}\
			\bibnamefont {Kippenberg}}, and\ \bibinfo {author} {\bibfnamefont
			{C.}~\bibnamefont {Koos}},\ }\bibfield  {title} {\bibinfo {title} {Ultrafast
			optical ranging using microresonator soliton frequency combs},\ }\href
	{https://doi.org/10.1126/science.aao3924} {\bibfield  {journal} {\bibinfo
			{journal} {Science}\ }\textbf {\bibinfo {volume} {359}},\ \bibinfo {pages}
		{887} (\bibinfo {year} {2018})}\BibitemShut {NoStop}%
	\bibitem [{\citenamefont {Suh}\ and\ \citenamefont
		{Vahala}(2018)}]{Suh_Science_2018}%
	\BibitemOpen
	\bibfield  {author} {\bibinfo {author} {\bibfnamefont {M.-G.}\ \bibnamefont
			{Suh}}and\ \bibinfo {author} {\bibfnamefont {K.~J.}\ \bibnamefont {Vahala}},\
	}\bibfield  {title} {\bibinfo {title} {Soliton microcomb range measurement},\
	}\href {https://doi.org/10.1126/science.aao1968} {\bibfield  {journal}
		{\bibinfo  {journal} {Science}\ }\textbf {\bibinfo {volume} {359}},\ \bibinfo
		{pages} {884} (\bibinfo {year} {2018})}\BibitemShut {NoStop}%
	\bibitem [{\citenamefont {Spencer}\ \emph {et~al.}(2018)\citenamefont
		{Spencer}, \citenamefont {Drake}, \citenamefont {Briles}, \citenamefont
		{Stone}, \citenamefont {Sinclair}, \citenamefont {Fredrick}, \citenamefont
		{Li}, \citenamefont {Westly}, \citenamefont {Ilic}, \citenamefont
		{Bluestone}, \citenamefont {Volet}, \citenamefont {Komljenovic},
		\citenamefont {Chang}, \citenamefont {Lee}, \citenamefont {Oh}, \citenamefont
		{Suh}, \citenamefont {Yang}, \citenamefont {Pfeiffer}, \citenamefont
		{Kippenberg}, \citenamefont {Norberg}, \citenamefont {Theogarajan},
		\citenamefont {Vahala}, \citenamefont {Newbury}, \citenamefont {Srinivasan},
		\citenamefont {Bowers}, \citenamefont {Diddams},\ and\ \citenamefont
		{Papp}}]{Spencer_Nature_2018}%
	\BibitemOpen
	\bibfield  {author} {\bibinfo {author} {\bibfnamefont {D.~T.}\ \bibnamefont
			{Spencer}}, \bibinfo {author} {\bibfnamefont {T.}~\bibnamefont {Drake}},
		\bibinfo {author} {\bibfnamefont {T.~C.}\ \bibnamefont {Briles}}, \bibinfo
		{author} {\bibfnamefont {J.}~\bibnamefont {Stone}}, \bibinfo {author}
		{\bibfnamefont {L.~C.}\ \bibnamefont {Sinclair}}, \bibinfo {author}
		{\bibfnamefont {C.}~\bibnamefont {Fredrick}}, \bibinfo {author}
		{\bibfnamefont {Q.}~\bibnamefont {Li}}, \bibinfo {author} {\bibfnamefont
			{D.}~\bibnamefont {Westly}}, \bibinfo {author} {\bibfnamefont {B.~R.}\
			\bibnamefont {Ilic}}, \bibinfo {author} {\bibfnamefont {A.}~\bibnamefont
			{Bluestone}}, \bibinfo {author} {\bibfnamefont {N.}~\bibnamefont {Volet}},
		\bibinfo {author} {\bibfnamefont {T.}~\bibnamefont {Komljenovic}}, \bibinfo
		{author} {\bibfnamefont {L.}~\bibnamefont {Chang}}, \bibinfo {author}
		{\bibfnamefont {S.~H.}\ \bibnamefont {Lee}}, \bibinfo {author} {\bibfnamefont
			{D.~Y.}\ \bibnamefont {Oh}}, \bibinfo {author} {\bibfnamefont {M.-G.}\
			\bibnamefont {Suh}}, \bibinfo {author} {\bibfnamefont {K.~Y.}\ \bibnamefont
			{Yang}}, \bibinfo {author} {\bibfnamefont {M.~H.~P.}\ \bibnamefont
			{Pfeiffer}}, \bibinfo {author} {\bibfnamefont {T.~J.}\ \bibnamefont
			{Kippenberg}}, \bibinfo {author} {\bibfnamefont {E.}~\bibnamefont {Norberg}},
		\bibinfo {author} {\bibfnamefont {L.}~\bibnamefont {Theogarajan}}, \bibinfo
		{author} {\bibfnamefont {K.}~\bibnamefont {Vahala}}, \bibinfo {author}
		{\bibfnamefont {N.~R.}\ \bibnamefont {Newbury}}, \bibinfo {author}
		{\bibfnamefont {K.}~\bibnamefont {Srinivasan}}, \bibinfo {author}
		{\bibfnamefont {J.~E.}\ \bibnamefont {Bowers}}, \bibinfo {author}
		{\bibfnamefont {S.~A.}\ \bibnamefont {Diddams}}, and\ \bibinfo {author}
		{\bibfnamefont {S.~B.}\ \bibnamefont {Papp}},\ }\bibfield  {title} {\bibinfo
		{title} {An optical-frequency synthesizer using integrated photonics},\
	}\href {https://doi.org/10.1038/s41586-018-0065-7} {\bibfield  {journal}
		{\bibinfo  {journal} {Nature}\ }\textbf {\bibinfo {volume} {557}},\ \bibinfo
		{pages} {81} (\bibinfo {year} {2018})}\BibitemShut {NoStop}%
	\bibitem [{\citenamefont {Marin-Palomo}\ \emph {et~al.}(2017)\citenamefont
		{Marin-Palomo}, \citenamefont {Kemal}, \citenamefont {Karpov}, \citenamefont
		{Kordts}, \citenamefont {Pfeifle}, \citenamefont {Pfeiffer}, \citenamefont
		{Trocha}, \citenamefont {Wolf}, \citenamefont {Brasch}, \citenamefont
		{Anderson}, \citenamefont {Rosenberger}, \citenamefont {Vijayan},
		\citenamefont {Freude}, \citenamefont {Kippenberg},\ and\ \citenamefont
		{Koos}}]{MarinPalomo_Nature_2017}%
	\BibitemOpen
	\bibfield  {author} {\bibinfo {author} {\bibfnamefont {P.}~\bibnamefont
			{Marin-Palomo}}, \bibinfo {author} {\bibfnamefont {J.~N.}\ \bibnamefont
			{Kemal}}, \bibinfo {author} {\bibfnamefont {M.}~\bibnamefont {Karpov}},
		\bibinfo {author} {\bibfnamefont {A.}~\bibnamefont {Kordts}}, \bibinfo
		{author} {\bibfnamefont {J.}~\bibnamefont {Pfeifle}}, \bibinfo {author}
		{\bibfnamefont {M.~H.~P.}\ \bibnamefont {Pfeiffer}}, \bibinfo {author}
		{\bibfnamefont {P.}~\bibnamefont {Trocha}}, \bibinfo {author} {\bibfnamefont
			{S.}~\bibnamefont {Wolf}}, \bibinfo {author} {\bibfnamefont {V.}~\bibnamefont
			{Brasch}}, \bibinfo {author} {\bibfnamefont {M.~H.}\ \bibnamefont
			{Anderson}}, \bibinfo {author} {\bibfnamefont {R.}~\bibnamefont
			{Rosenberger}}, \bibinfo {author} {\bibfnamefont {K.}~\bibnamefont
			{Vijayan}}, \bibinfo {author} {\bibfnamefont {W.}~\bibnamefont {Freude}},
		\bibinfo {author} {\bibfnamefont {T.~J.}\ \bibnamefont {Kippenberg}}, and\
		\bibinfo {author} {\bibfnamefont {C.}~\bibnamefont {Koos}},\ }\bibfield
	{title} {\bibinfo {title} {Microresonator-based solitons for massively
			parallel coherent optical communications},\ }\href
	{https://doi.org/10.1038/nature22387} {\bibfield  {journal} {\bibinfo
			{journal} {Nature}\ }\textbf {\bibinfo {volume} {546}},\ \bibinfo {pages}
		{274} (\bibinfo {year} {2017})}\BibitemShut {NoStop}%
	\bibitem [{\citenamefont {Vanier}(2005)}]{Vanier_ApplPhysB_2005}%
	\BibitemOpen
	\bibfield  {author} {\bibinfo {author} {\bibfnamefont {J.}~\bibnamefont
			{Vanier}},\ }\bibfield  {title} {\bibinfo {title} {Atomic clocks based on
			coherent population trapping: a review},\ }\href
	{https://doi.org/10.1007/s00340-005-1905-3} {\bibfield  {journal} {\bibinfo
			{journal} {Appl. Phys. B}\ }\textbf {\bibinfo {volume} {81}},\ \bibinfo
		{pages} {421} (\bibinfo {year} {2005})}\BibitemShut {NoStop}%
	\bibitem [{\citenamefont {Papp}\ \emph {et~al.}(2014)\citenamefont {Papp},
		\citenamefont {Beha}, \citenamefont {Del'Haye}, \citenamefont {Quinlan},
		\citenamefont {Lee}, \citenamefont {Vahala},\ and\ \citenamefont
		{Diddams}}]{Papp_Optica_2014}%
	\BibitemOpen
	\bibfield  {author} {\bibinfo {author} {\bibfnamefont {S.~B.}\ \bibnamefont
			{Papp}}, \bibinfo {author} {\bibfnamefont {K.}~\bibnamefont {Beha}}, \bibinfo
		{author} {\bibfnamefont {P.}~\bibnamefont {Del'Haye}}, \bibinfo {author}
		{\bibfnamefont {F.}~\bibnamefont {Quinlan}}, \bibinfo {author} {\bibfnamefont
			{H.}~\bibnamefont {Lee}}, \bibinfo {author} {\bibfnamefont {K.~J.}\
			\bibnamefont {Vahala}}, and\ \bibinfo {author} {\bibfnamefont {S.~A.}\
			\bibnamefont {Diddams}},\ }\bibfield  {title} {\bibinfo {title}
		{Microresonator frequency comb optical clock},\ }\href
	{https://doi.org/10.1364/OPTICA.1.000010} {\bibfield  {journal} {\bibinfo
			{journal} {Optica}\ }\textbf {\bibinfo {volume} {1}},\ \bibinfo {pages} {10}
		(\bibinfo {year} {2014})}\BibitemShut {NoStop}%
	\bibitem [{\citenamefont {Li}\ \emph {et~al.}(2008)\citenamefont {Li},
		\citenamefont {Benedick}, \citenamefont {Fendel}, \citenamefont {Glenday},
		\citenamefont {Kärtner}, \citenamefont {Phillips}, \citenamefont {Sasselov},
		\citenamefont {Szentgyorgyi},\ and\ \citenamefont
		{Walsworth}}]{Li_Nature_2008}%
	\BibitemOpen
	\bibfield  {author} {\bibinfo {author} {\bibfnamefont {C.-H.}\ \bibnamefont
			{Li}}, \bibinfo {author} {\bibfnamefont {A.~J.}\ \bibnamefont {Benedick}},
		\bibinfo {author} {\bibfnamefont {P.}~\bibnamefont {Fendel}}, \bibinfo
		{author} {\bibfnamefont {A.~G.}\ \bibnamefont {Glenday}}, \bibinfo {author}
		{\bibfnamefont {F.~X.}\ \bibnamefont {Kärtner}}, \bibinfo {author}
		{\bibfnamefont {D.~F.}\ \bibnamefont {Phillips}}, \bibinfo {author}
		{\bibfnamefont {D.}~\bibnamefont {Sasselov}}, \bibinfo {author}
		{\bibfnamefont {A.}~\bibnamefont {Szentgyorgyi}}, and\ \bibinfo {author}
		{\bibfnamefont {R.~L.}\ \bibnamefont {Walsworth}},\ }\bibfield  {title}
	{\bibinfo {title} {A laser frequency comb that enables radial velocity
			measurements with a precision of 1 cm s$^{-1}$},\ }\href
	{https://doi.org/10.1038/nature06854} {\bibfield  {journal} {\bibinfo
			{journal} {Nature}\ }\textbf {\bibinfo {volume} {452}},\ \bibinfo {pages}
		{610} (\bibinfo {year} {2008})}\BibitemShut {NoStop}%
	\bibitem [{\citenamefont {Obrzud}\ \emph {et~al.}(2019)\citenamefont {Obrzud},
		\citenamefont {Rainer}, \citenamefont {Harutyunyan}, \citenamefont
		{Anderson}, \citenamefont {Liu}, \citenamefont {Geiselmann}, \citenamefont
		{Chazelas}, \citenamefont {Kundermann}, \citenamefont {Lecomte},
		\citenamefont {Cecconi}, \citenamefont {Ghedina}, \citenamefont {Molinari},
		\citenamefont {Pepe}, \citenamefont {Wildi}, \citenamefont {Bouchy},
		\citenamefont {Kippenberg},\ and\ \citenamefont
		{Herr}}]{Obrzud_NatPhot_2019}%
	\BibitemOpen
	\bibfield  {author} {\bibinfo {author} {\bibfnamefont {E.}~\bibnamefont
			{Obrzud}}, \bibinfo {author} {\bibfnamefont {M.}~\bibnamefont {Rainer}},
		\bibinfo {author} {\bibfnamefont {A.}~\bibnamefont {Harutyunyan}}, \bibinfo
		{author} {\bibfnamefont {M.~H.}\ \bibnamefont {Anderson}}, \bibinfo {author}
		{\bibfnamefont {J.}~\bibnamefont {Liu}}, \bibinfo {author} {\bibfnamefont
			{M.}~\bibnamefont {Geiselmann}}, \bibinfo {author} {\bibfnamefont
			{B.}~\bibnamefont {Chazelas}}, \bibinfo {author} {\bibfnamefont
			{S.}~\bibnamefont {Kundermann}}, \bibinfo {author} {\bibfnamefont
			{S.}~\bibnamefont {Lecomte}}, \bibinfo {author} {\bibfnamefont
			{M.}~\bibnamefont {Cecconi}}, \bibinfo {author} {\bibfnamefont
			{A.}~\bibnamefont {Ghedina}}, \bibinfo {author} {\bibfnamefont
			{E.}~\bibnamefont {Molinari}}, \bibinfo {author} {\bibfnamefont
			{F.}~\bibnamefont {Pepe}}, \bibinfo {author} {\bibfnamefont {F.}~\bibnamefont
			{Wildi}}, \bibinfo {author} {\bibfnamefont {F.}~\bibnamefont {Bouchy}},
		\bibinfo {author} {\bibfnamefont {T.~J.}\ \bibnamefont {Kippenberg}}, and\
		\bibinfo {author} {\bibfnamefont {T.}~\bibnamefont {Herr}},\ }\bibfield
	{title} {\bibinfo {title} {A microphotonic astrocomb},\ }\href
	{https://doi.org/10.1038/s41566-018-0309-y} {\bibfield  {journal} {\bibinfo
			{journal} {Nat. Photonics}\ }\textbf {\bibinfo {volume} {13}},\ \bibinfo
		{pages} {31} (\bibinfo {year} {2019})}\BibitemShut {NoStop}%
	\bibitem [{\citenamefont {Suh}\ \emph {et~al.}(2019)\citenamefont {Suh},
		\citenamefont {Yi}, \citenamefont {Lai}, \citenamefont {Leifer},
		\citenamefont {Grudinin}, \citenamefont {Vasisht}, \citenamefont {Martin},
		\citenamefont {Fitzgerald}, \citenamefont {Doppmann}, \citenamefont {Wang},
		\citenamefont {Mawet}, \citenamefont {Papp}, \citenamefont {Diddams},
		\citenamefont {Beichman},\ and\ \citenamefont {Vahala}}]{Suh_NatPhot_2019}%
	\BibitemOpen
	\bibfield  {author} {\bibinfo {author} {\bibfnamefont {M.-G.}\ \bibnamefont
			{Suh}}, \bibinfo {author} {\bibfnamefont {X.}~\bibnamefont {Yi}}, \bibinfo
		{author} {\bibfnamefont {Y.-H.}\ \bibnamefont {Lai}}, \bibinfo {author}
		{\bibfnamefont {S.}~\bibnamefont {Leifer}}, \bibinfo {author} {\bibfnamefont
			{I.~S.}\ \bibnamefont {Grudinin}}, \bibinfo {author} {\bibfnamefont
			{G.}~\bibnamefont {Vasisht}}, \bibinfo {author} {\bibfnamefont {E.~C.}\
			\bibnamefont {Martin}}, \bibinfo {author} {\bibfnamefont {M.~P.}\
			\bibnamefont {Fitzgerald}}, \bibinfo {author} {\bibfnamefont
			{G.}~\bibnamefont {Doppmann}}, \bibinfo {author} {\bibfnamefont
			{J.}~\bibnamefont {Wang}}, \bibinfo {author} {\bibfnamefont {D.}~\bibnamefont
			{Mawet}}, \bibinfo {author} {\bibfnamefont {S.~B.}\ \bibnamefont {Papp}},
		\bibinfo {author} {\bibfnamefont {S.~A.}\ \bibnamefont {Diddams}}, \bibinfo
		{author} {\bibfnamefont {C.}~\bibnamefont {Beichman}}, and\ \bibinfo {author}
		{\bibfnamefont {K.}~\bibnamefont {Vahala}},\ }\bibfield  {title} {\bibinfo
		{title} {Searching for exoplanets using a microresonator astrocomb},\ }\href
	{https://doi.org/10.1038/s41566-018-0312-3} {\bibfield  {journal} {\bibinfo
			{journal} {Nat. Photonics}\ }\textbf {\bibinfo {volume} {13}},\ \bibinfo
		{pages} {25} (\bibinfo {year} {2019})}\BibitemShut {NoStop}%
	\bibitem [{\citenamefont {Lee}\ \emph {et~al.}(2001)\citenamefont {Lee},
		\citenamefont {Widiyatmoko}, \citenamefont {Kourogi},\ and\ \citenamefont
		{Ohtsu}}]{Lee_Misc_2001}%
	\BibitemOpen
	\bibfield  {author} {\bibinfo {author} {\bibfnamefont {S.-J.}\ \bibnamefont
			{Lee}}, \bibinfo {author} {\bibfnamefont {B.}~\bibnamefont {Widiyatmoko}},
		\bibinfo {author} {\bibfnamefont {M.}~\bibnamefont {Kourogi}}, and\ \bibinfo
		{author} {\bibfnamefont {M.}~\bibnamefont {Ohtsu}},\ }\bibfield  {title}
	{\bibinfo {title} {Ultrahigh scanning speed optical coherence tomography
			using optical frequency comb generators},\ }\href
	{http://stacks.iop.org/1347-4065/40/i=8B/a=L878} {\bibfield  {journal}
		{\bibinfo  {journal} {Jpn. J. Appl. Phys.}\ }\textbf {\bibinfo {volume}
			{40}},\ \bibinfo {pages} {L878} (\bibinfo {year} {2001})}\BibitemShut
	{NoStop}%
	\bibitem [{\citenamefont {Fercher}\ \emph {et~al.}(2003)\citenamefont
		{Fercher}, \citenamefont {Drexler}, \citenamefont {Hitzenberger},\ and\
		\citenamefont {Lasser}}]{Fercher_ReportsProgressPhys_2003}%
	\BibitemOpen
	\bibfield  {author} {\bibinfo {author} {\bibfnamefont {A.~F.}\ \bibnamefont
			{Fercher}}, \bibinfo {author} {\bibfnamefont {W.}~\bibnamefont {Drexler}},
		\bibinfo {author} {\bibfnamefont {C.~K.}\ \bibnamefont {Hitzenberger}}, and\
		\bibinfo {author} {\bibfnamefont {T.}~\bibnamefont {Lasser}},\ }\bibfield
	{title} {\bibinfo {title} {Optical coherence tomography - principles and
			applications},\ }\href {http://stacks.iop.org/0034-4885/66/i=2/a=204}
	{\bibfield  {journal} {\bibinfo  {journal} {Rep. Prog. Phys.}\ }\textbf
		{\bibinfo {volume} {66}},\ \bibinfo {pages} {239} (\bibinfo {year}
		{2003})}\BibitemShut {NoStop}%
	\bibitem [{\citenamefont {Saha}\ \emph {et~al.}(2012)\citenamefont {Saha},
		\citenamefont {Okawachi}, \citenamefont {Levy}, \citenamefont {Lau},
		\citenamefont {Luke}, \citenamefont {Foster}, \citenamefont {Lipson},\ and\
		\citenamefont {Gaeta}}]{Saha_OE_2012}%
	\BibitemOpen
	\bibfield  {author} {\bibinfo {author} {\bibfnamefont {K.}~\bibnamefont
			{Saha}}, \bibinfo {author} {\bibfnamefont {Y.}~\bibnamefont {Okawachi}},
		\bibinfo {author} {\bibfnamefont {J.~S.}\ \bibnamefont {Levy}}, \bibinfo
		{author} {\bibfnamefont {R.~K.~W.}\ \bibnamefont {Lau}}, \bibinfo {author}
		{\bibfnamefont {K.}~\bibnamefont {Luke}}, \bibinfo {author} {\bibfnamefont
			{M.~A.}\ \bibnamefont {Foster}}, \bibinfo {author} {\bibfnamefont
			{M.}~\bibnamefont {Lipson}}, and\ \bibinfo {author} {\bibfnamefont {A.~L.}\
			\bibnamefont {Gaeta}},\ }\bibfield  {title} {\bibinfo {title} {Broadband
			parametric frequency comb generation with a 1-$\mu$m pump source},\ }\href
	{https://doi.org/10.1364/OE.20.026935} {\bibfield  {journal} {\bibinfo
			{journal} {Opt. Express}\ }\textbf {\bibinfo {volume} {20}},\ \bibinfo
		{pages} {26935} (\bibinfo {year} {2012})}\BibitemShut {NoStop}%
	\bibitem [{\citenamefont {Yang}\ \emph {et~al.}(2016)\citenamefont {Yang},
		\citenamefont {Jiang}, \citenamefont {Kasumie}, \citenamefont {Zhao},
		\citenamefont {Xu}, \citenamefont {Ward}, \citenamefont {Yang},\ and\
		\citenamefont {Chormaic}}]{Yang_OL_2016}%
	\BibitemOpen
	\bibfield  {author} {\bibinfo {author} {\bibfnamefont {Y.}~\bibnamefont
			{Yang}}, \bibinfo {author} {\bibfnamefont {X.}~\bibnamefont {Jiang}},
		\bibinfo {author} {\bibfnamefont {S.}~\bibnamefont {Kasumie}}, \bibinfo
		{author} {\bibfnamefont {G.}~\bibnamefont {Zhao}}, \bibinfo {author}
		{\bibfnamefont {L.}~\bibnamefont {Xu}}, \bibinfo {author} {\bibfnamefont
			{J.~M.}\ \bibnamefont {Ward}}, \bibinfo {author} {\bibfnamefont
			{L.}~\bibnamefont {Yang}}, and\ \bibinfo {author} {\bibfnamefont {S.~N.}\
			\bibnamefont {Chormaic}},\ }\bibfield  {title} {\bibinfo {title} {Four-wave
			mixing parametric oscillation and frequency comb generation at visible
			wavelengths in a silica microbubble resonator},\ }\href
	{https://doi.org/10.1364/OL.41.005266} {\bibfield  {journal} {\bibinfo
			{journal} {Opt. Lett.}\ }\textbf {\bibinfo {volume} {41}},\ \bibinfo {pages}
		{5266} (\bibinfo {year} {2016})}\BibitemShut {NoStop}%
	\bibitem [{\citenamefont {Wang}\ \emph {et~al.}(2016)\citenamefont {Wang},
		\citenamefont {Chang}, \citenamefont {Volet}, \citenamefont {Pfeiffer},
		\citenamefont {Zervas}, \citenamefont {Guo}, \citenamefont {Kippenberg},\
		and\ \citenamefont {Bowers}}]{Wang_LaserPhotRev_2016}%
	\BibitemOpen
	\bibfield  {author} {\bibinfo {author} {\bibfnamefont {L.}~\bibnamefont
			{Wang}}, \bibinfo {author} {\bibfnamefont {L.}~\bibnamefont {Chang}},
		\bibinfo {author} {\bibfnamefont {N.}~\bibnamefont {Volet}}, \bibinfo
		{author} {\bibfnamefont {M.~H.~P.}\ \bibnamefont {Pfeiffer}}, \bibinfo
		{author} {\bibfnamefont {M.}~\bibnamefont {Zervas}}, \bibinfo {author}
		{\bibfnamefont {H.}~\bibnamefont {Guo}}, \bibinfo {author} {\bibfnamefont
			{T.~J.}\ \bibnamefont {Kippenberg}}, and\ \bibinfo {author} {\bibfnamefont
			{J.~E.}\ \bibnamefont {Bowers}},\ }\bibfield  {title} {\bibinfo {title}
		{Frequency comb generation in the green using silicon nitride
			microresonators},\ }\href {https://doi.org/10.1002/lpor.201600006} {\bibfield
		{journal} {\bibinfo  {journal} {Laser Photonics Rev.}\ }\textbf {\bibinfo
			{volume} {10}},\ \bibinfo {pages} {631} (\bibinfo {year} {2016})}\BibitemShut
	{NoStop}%
	\bibitem [{\citenamefont {Soltani}\ \emph {et~al.}(2016)\citenamefont
		{Soltani}, \citenamefont {Matsko},\ and\ \citenamefont
		{Maleki}}]{Soltani_LaserPhotRev_2016}%
	\BibitemOpen
	\bibfield  {author} {\bibinfo {author} {\bibfnamefont {M.}~\bibnamefont
			{Soltani}}, \bibinfo {author} {\bibfnamefont {A.}~\bibnamefont {Matsko}},
		and\ \bibinfo {author} {\bibfnamefont {L.}~\bibnamefont {Maleki}},\
	}\bibfield  {title} {\bibinfo {title} {Enabling arbitrary wavelength
			frequency combs on chip},\ }\href {https://doi.org/10.1002/lpor.201500226}
	{\bibfield  {journal} {\bibinfo  {journal} {Laser Photonics Rev.}\ }\textbf
		{\bibinfo {volume} {10}},\ \bibinfo {pages} {158} (\bibinfo {year}
		{2016})}\BibitemShut {NoStop}%
	\bibitem [{\citenamefont {Miller}\ \emph {et~al.}(2014)\citenamefont {Miller},
		\citenamefont {Luke}, \citenamefont {Okawachi}, \citenamefont {Cardenas},
		\citenamefont {Gaeta},\ and\ \citenamefont {Lipson}}]{Miller_OE_2014}%
	\BibitemOpen
	\bibfield  {author} {\bibinfo {author} {\bibfnamefont {S.}~\bibnamefont
			{Miller}}, \bibinfo {author} {\bibfnamefont {K.}~\bibnamefont {Luke}},
		\bibinfo {author} {\bibfnamefont {Y.}~\bibnamefont {Okawachi}}, \bibinfo
		{author} {\bibfnamefont {J.}~\bibnamefont {Cardenas}}, \bibinfo {author}
		{\bibfnamefont {A.~L.}\ \bibnamefont {Gaeta}}, and\ \bibinfo {author}
		{\bibfnamefont {M.}~\bibnamefont {Lipson}},\ }\bibfield  {title} {\bibinfo
		{title} {On-chip frequency comb generation at visible wavelengths via
			simultaneous second- and third-order optical nonlinearities},\ }\href
	{https://doi.org/10.1364/OE.22.026517} {\bibfield  {journal} {\bibinfo
			{journal} {Opt. Express}\ }\textbf {\bibinfo {volume} {22}},\ \bibinfo
		{pages} {26517} (\bibinfo {year} {2014})}\BibitemShut {NoStop}%
	\bibitem [{\citenamefont {Guo}\ \emph {et~al.}(2018)\citenamefont {Guo},
		\citenamefont {Zou}, \citenamefont {Jung}, \citenamefont {Gong},
		\citenamefont {Bruch}, \citenamefont {Jiang},\ and\ \citenamefont
		{Tang}}]{Guo_PRAppl_2018}%
	\BibitemOpen
	\bibfield  {author} {\bibinfo {author} {\bibfnamefont {X.}~\bibnamefont
			{Guo}}, \bibinfo {author} {\bibfnamefont {C.-L.}\ \bibnamefont {Zou}},
		\bibinfo {author} {\bibfnamefont {H.}~\bibnamefont {Jung}}, \bibinfo {author}
		{\bibfnamefont {Z.}~\bibnamefont {Gong}}, \bibinfo {author} {\bibfnamefont
			{A.}~\bibnamefont {Bruch}}, \bibinfo {author} {\bibfnamefont
			{L.}~\bibnamefont {Jiang}}, and\ \bibinfo {author} {\bibfnamefont {H.~X.}\
			\bibnamefont {Tang}},\ }\bibfield  {title} {\bibinfo {title} {Efficient
			generation of a near-visible frequency comb via cherenkov-like radiation from
			a kerr microcomb},\ }\href {https://doi.org/10.1103/PhysRevApplied.10.014012}
	{\bibfield  {journal} {\bibinfo  {journal} {Phys. Rev. Appl.}\ }\textbf
		{\bibinfo {volume} {10}},\ \bibinfo {pages} {014012} (\bibinfo {year}
		{2018})}\BibitemShut {NoStop}%
	\bibitem [{\citenamefont {Lee}\ \emph {et~al.}(2017)\citenamefont {Lee},
		\citenamefont {Oh}, \citenamefont {Yang}, \citenamefont {Shen}, \citenamefont
		{Wang}, \citenamefont {Yang}, \citenamefont {Lai}, \citenamefont {Yi},
		\citenamefont {Li},\ and\ \citenamefont {Vahala}}]{Lee_NatComm_2017}%
	\BibitemOpen
	\bibfield  {author} {\bibinfo {author} {\bibfnamefont {S.~H.}\ \bibnamefont
			{Lee}}, \bibinfo {author} {\bibfnamefont {D.~Y.}\ \bibnamefont {Oh}},
		\bibinfo {author} {\bibfnamefont {Q.-F.}\ \bibnamefont {Yang}}, \bibinfo
		{author} {\bibfnamefont {B.}~\bibnamefont {Shen}}, \bibinfo {author}
		{\bibfnamefont {H.}~\bibnamefont {Wang}}, \bibinfo {author} {\bibfnamefont
			{K.~Y.}\ \bibnamefont {Yang}}, \bibinfo {author} {\bibfnamefont {Y.-H.}\
			\bibnamefont {Lai}}, \bibinfo {author} {\bibfnamefont {X.}~\bibnamefont
			{Yi}}, \bibinfo {author} {\bibfnamefont {X.}~\bibnamefont {Li}}, and\
		\bibinfo {author} {\bibfnamefont {K.}~\bibnamefont {Vahala}},\ }\bibfield
	{title} {\bibinfo {title} {Towards visible soliton microcomb generation},\
	}\href {https://doi.org/10.1038/s41467-017-01473-9} {\bibfield  {journal}
		{\bibinfo  {journal} {Nat. Commun.}\ }\textbf {\bibinfo {volume} {8}},\
		\bibinfo {pages} {1295} (\bibinfo {year} {2017})}\BibitemShut {NoStop}%
	\bibitem [{\citenamefont {Yu}\ \emph {et~al.}(2019)\citenamefont {Yu},
		\citenamefont {Briles}, \citenamefont {Moille}, \citenamefont {Lu},
		\citenamefont {Diddams}, \citenamefont {Srinivasan},\ and\ \citenamefont
		{Papp}}]{Yu_PRAppl_2019}%
	\BibitemOpen
	\bibfield  {author} {\bibinfo {author} {\bibfnamefont {S.-P.}\ \bibnamefont
			{Yu}}, \bibinfo {author} {\bibfnamefont {T.~C.}\ \bibnamefont {Briles}},
		\bibinfo {author} {\bibfnamefont {G.~T.}\ \bibnamefont {Moille}}, \bibinfo
		{author} {\bibfnamefont {X.}~\bibnamefont {Lu}}, \bibinfo {author}
		{\bibfnamefont {S.~A.}\ \bibnamefont {Diddams}}, \bibinfo {author}
		{\bibfnamefont {K.}~\bibnamefont {Srinivasan}}, and\ \bibinfo {author}
		{\bibfnamefont {S.~B.}\ \bibnamefont {Papp}},\ }\bibfield  {title} {\bibinfo
		{title} {Tuning kerr-soliton frequency combs to atomic resonances},\ }\href
	{https://doi.org/10.1103/PhysRevApplied.11.044017} {\bibfield  {journal}
		{\bibinfo  {journal} {Phys. Rev. Appl.}\ }\textbf {\bibinfo {volume} {11}},\
		\bibinfo {pages} {044017} (\bibinfo {year} {2019})}\BibitemShut {NoStop}%
	\bibitem [{\citenamefont {Joshi}\ \emph {et~al.}(2016)\citenamefont {Joshi},
		\citenamefont {Jang}, \citenamefont {Luke}, \citenamefont {Ji}, \citenamefont
		{Miller}, \citenamefont {Klenner}, \citenamefont {Okawachi}, \citenamefont
		{Lipson},\ and\ \citenamefont {Gaeta}}]{Joshi_OL_2016}%
	\BibitemOpen
	\bibfield  {author} {\bibinfo {author} {\bibfnamefont {C.}~\bibnamefont
			{Joshi}}, \bibinfo {author} {\bibfnamefont {J.~K.}\ \bibnamefont {Jang}},
		\bibinfo {author} {\bibfnamefont {K.}~\bibnamefont {Luke}}, \bibinfo {author}
		{\bibfnamefont {X.}~\bibnamefont {Ji}}, \bibinfo {author} {\bibfnamefont
			{S.~A.}\ \bibnamefont {Miller}}, \bibinfo {author} {\bibfnamefont
			{A.}~\bibnamefont {Klenner}}, \bibinfo {author} {\bibfnamefont
			{Y.}~\bibnamefont {Okawachi}}, \bibinfo {author} {\bibfnamefont
			{M.}~\bibnamefont {Lipson}}, and\ \bibinfo {author} {\bibfnamefont {A.~L.}\
			\bibnamefont {Gaeta}},\ }\bibfield  {title} {\bibinfo {title} {Thermally
			controlled comb generation and soliton modelocking in microresonators},\
	}\href {https://doi.org/10.1364/OL.41.002565} {\bibfield  {journal} {\bibinfo
			{journal} {Opt. Lett.}\ }\textbf {\bibinfo {volume} {41}},\ \bibinfo {pages}
		{2565} (\bibinfo {year} {2016})}\BibitemShut {NoStop}%
	\bibitem [{\citenamefont {Cole}\ \emph {et~al.}(2017)\citenamefont {Cole},
		\citenamefont {Lamb}, \citenamefont {Del’Haye}, \citenamefont {Diddams},\
		and\ \citenamefont {Papp}}]{cole_NatPhot_2017}%
	\BibitemOpen
	\bibfield  {author} {\bibinfo {author} {\bibfnamefont {D.~C.}\ \bibnamefont
			{Cole}}, \bibinfo {author} {\bibfnamefont {E.~S.}\ \bibnamefont {Lamb}},
		\bibinfo {author} {\bibfnamefont {P.}~\bibnamefont {Del’Haye}}, \bibinfo
		{author} {\bibfnamefont {S.~A.}\ \bibnamefont {Diddams}}, and\ \bibinfo
		{author} {\bibfnamefont {S.~B.}\ \bibnamefont {Papp}},\ }\bibfield  {title}
	{\bibinfo {title} {Soliton crystals in {Kerr} resonators},\ }\href
	{https://doi.org/10.1038/s41566-017-0009-z} {\bibfield  {journal} {\bibinfo
			{journal} {Nat. Photonics}\ }\textbf {\bibinfo {volume} {11}},\ \bibinfo
		{pages} {671} (\bibinfo {year} {2017})}\BibitemShut {NoStop}%
	\bibitem [{\citenamefont {Carmon}\ \emph {et~al.}(2004)\citenamefont {Carmon},
		\citenamefont {Yang},\ and\ \citenamefont {Vahala}}]{Carmon_OE_2004}%
	\BibitemOpen
	\bibfield  {author} {\bibinfo {author} {\bibfnamefont {T.}~\bibnamefont
			{Carmon}}, \bibinfo {author} {\bibfnamefont {L.}~\bibnamefont {Yang}}, and\
		\bibinfo {author} {\bibfnamefont {K.~J.}\ \bibnamefont {Vahala}},\ }\bibfield
	{title} {\bibinfo {title} {Dynamical thermal behavior and thermal
			self-stability of microcavities},\ }\href
	{https://doi.org/10.1364/OPEX.12.004742} {\bibfield  {journal} {\bibinfo
			{journal} {Opt. Express}\ }\textbf {\bibinfo {volume} {12}},\ \bibinfo
		{pages} {4742} (\bibinfo {year} {2004})}\BibitemShut {NoStop}%
	\bibitem [{\citenamefont {Siddons}\ \emph {et~al.}(2008)\citenamefont
		{Siddons}, \citenamefont {Adams}, \citenamefont {Ge},\ and\ \citenamefont
		{Hughes}}]{Siddons_JPhys_2008}%
	\BibitemOpen
	\bibfield  {author} {\bibinfo {author} {\bibfnamefont {P.}~\bibnamefont
			{Siddons}}, \bibinfo {author} {\bibfnamefont {C.~S.}\ \bibnamefont {Adams}},
		\bibinfo {author} {\bibfnamefont {C.}~\bibnamefont {Ge}}, and\ \bibinfo
		{author} {\bibfnamefont {I.~G.}\ \bibnamefont {Hughes}},\ }\bibfield  {title}
	{\bibinfo {title} {Absolute absorption on rubidium d lines: comparison
			between theory and experiment},\ }\href
	{https://doi.org/10.1088/0953-4075/41/15/155004} {\bibfield  {journal}
		{\bibinfo  {journal} {J. Phys. B}\ }\textbf {\bibinfo {volume} {41}},\
		\bibinfo {pages} {155004} (\bibinfo {year} {2008})}\BibitemShut {NoStop}%
	\bibitem [{\citenamefont {Liu}\ \emph {et~al.}(2001)\citenamefont {Liu},
		\citenamefont {Dutton}, \citenamefont {Behroozi},\ and\ \citenamefont
		{Hau}}]{Liu_Nature_2001}%
	\BibitemOpen
	\bibfield  {author} {\bibinfo {author} {\bibfnamefont {C.}~\bibnamefont
			{Liu}}, \bibinfo {author} {\bibfnamefont {Z.}~\bibnamefont {Dutton}},
		\bibinfo {author} {\bibfnamefont {C.~H.}\ \bibnamefont {Behroozi}}, and\
		\bibinfo {author} {\bibfnamefont {L.~V.}\ \bibnamefont {Hau}},\ }\bibfield
	{title} {\bibinfo {title} {Observation of coherent optical information
			storage in an atomic medium using halted light pulses},\ }\href
	{https://doi.org/10.1038/35054017} {\bibfield  {journal} {\bibinfo  {journal}
			{Nature}\ }\textbf {\bibinfo {volume} {409}},\ \bibinfo {pages} {490}
		(\bibinfo {year} {2001})}\BibitemShut {NoStop}%
	\bibitem [{\citenamefont {Phillips}\ \emph {et~al.}(2001)\citenamefont
		{Phillips}, \citenamefont {Fleischhauer}, \citenamefont {Mair}, \citenamefont
		{Walsworth},\ and\ \citenamefont {Lukin}}]{Phillips_PRL_2001}%
	\BibitemOpen
	\bibfield  {author} {\bibinfo {author} {\bibfnamefont {D.~F.}\ \bibnamefont
			{Phillips}}, \bibinfo {author} {\bibfnamefont {A.}~\bibnamefont
			{Fleischhauer}}, \bibinfo {author} {\bibfnamefont {A.}~\bibnamefont {Mair}},
		\bibinfo {author} {\bibfnamefont {R.~L.}\ \bibnamefont {Walsworth}}, and\
		\bibinfo {author} {\bibfnamefont {M.~D.}\ \bibnamefont {Lukin}},\ }\bibfield
	{title} {\bibinfo {title} {Storage of light in atomic vapor},\ }\href
	{https://doi.org/10.1103/PhysRevLett.86.783} {\bibfield  {journal} {\bibinfo
			{journal} {Phys. Rev. Lett.}\ }\textbf {\bibinfo {volume} {86}},\ \bibinfo
		{pages} {783} (\bibinfo {year} {2001})}\BibitemShut {NoStop}%
	\bibitem [{\citenamefont {Chanelière}\ \emph {et~al.}(2005)\citenamefont
		{Chanelière}, \citenamefont {Matsukevich}, \citenamefont {Jenkins},
		\citenamefont {Lan}, \citenamefont {Kennedy},\ and\ \citenamefont
		{Kuzmich}}]{Chaneliere_Nature_2005}%
	\BibitemOpen
	\bibfield  {author} {\bibinfo {author} {\bibfnamefont {T.}~\bibnamefont
			{Chanelière}}, \bibinfo {author} {\bibfnamefont {D.~N.}\ \bibnamefont
			{Matsukevich}}, \bibinfo {author} {\bibfnamefont {S.~D.}\ \bibnamefont
			{Jenkins}}, \bibinfo {author} {\bibfnamefont {S.-Y.}\ \bibnamefont {Lan}},
		\bibinfo {author} {\bibfnamefont {T.~A.~B.}\ \bibnamefont {Kennedy}}, and\
		\bibinfo {author} {\bibfnamefont {A.}~\bibnamefont {Kuzmich}},\ }\bibfield
	{title} {\bibinfo {title} {Storage and retrieval of single photons
			transmitted between remote quantum memories},\ }\href
	{https://doi.org/10.1038/nature04315} {\bibfield  {journal} {\bibinfo
			{journal} {Nature}\ }\textbf {\bibinfo {volume} {438}},\ \bibinfo {pages}
		{833} (\bibinfo {year} {2005})}\BibitemShut {NoStop}%
	\bibitem [{\citenamefont {Bali\ifmmode~\acute{c}\else \'{c}\fi{}}\ \emph
		{et~al.}(2005)\citenamefont {Bali\ifmmode~\acute{c}\else \'{c}\fi{}},
		\citenamefont {Braje}, \citenamefont {Kolchin}, \citenamefont {Yin},\ and\
		\citenamefont {Harris}}]{Balic_PRL_2005}%
	\BibitemOpen
	\bibfield  {author} {\bibinfo {author} {\bibfnamefont {V.}~\bibnamefont
			{Bali\ifmmode~\acute{c}\else \'{c}\fi{}}}, \bibinfo {author} {\bibfnamefont
			{D.~A.}\ \bibnamefont {Braje}}, \bibinfo {author} {\bibfnamefont
			{P.}~\bibnamefont {Kolchin}}, \bibinfo {author} {\bibfnamefont {G.~Y.}\
			\bibnamefont {Yin}}, and\ \bibinfo {author} {\bibfnamefont {S.~E.}\
			\bibnamefont {Harris}},\ }\bibfield  {title} {\bibinfo {title} {Generation of
			paired photons with controllable waveforms},\ }\href
	{https://doi.org/10.1103/PhysRevLett.94.183601} {\bibfield  {journal}
		{\bibinfo  {journal} {Phys. Rev. Lett.}\ }\textbf {\bibinfo {volume} {94}},\
		\bibinfo {pages} {183601} (\bibinfo {year} {2005})}\BibitemShut {NoStop}%
	\bibitem [{\citenamefont {Förtsch}\ \emph {et~al.}(2013)\citenamefont
		{Förtsch}, \citenamefont {Fürst}, \citenamefont {Wittmann}, \citenamefont
		{Strekalov}, \citenamefont {Aiello}, \citenamefont {Chekhova}, \citenamefont
		{Silberhorn}, \citenamefont {Leuchs},\ and\ \citenamefont
		{Marquardt}}]{Fortsch_NatComm_2013}%
	\BibitemOpen
	\bibfield  {author} {\bibinfo {author} {\bibfnamefont {M.}~\bibnamefont
			{Förtsch}}, \bibinfo {author} {\bibfnamefont {J.~U.}\ \bibnamefont
			{Fürst}}, \bibinfo {author} {\bibfnamefont {C.}~\bibnamefont {Wittmann}},
		\bibinfo {author} {\bibfnamefont {D.}~\bibnamefont {Strekalov}}, \bibinfo
		{author} {\bibfnamefont {A.}~\bibnamefont {Aiello}}, \bibinfo {author}
		{\bibfnamefont {M.~V.}\ \bibnamefont {Chekhova}}, \bibinfo {author}
		{\bibfnamefont {C.}~\bibnamefont {Silberhorn}}, \bibinfo {author}
		{\bibfnamefont {G.}~\bibnamefont {Leuchs}}, and\ \bibinfo {author}
		{\bibfnamefont {C.}~\bibnamefont {Marquardt}},\ }\bibfield  {title} {\bibinfo
		{title} {A versatile source of single photons for quantum information
			processing},\ }\href {https://doi.org/10.1038/ncomms2838} {\bibfield
		{journal} {\bibinfo  {journal} {Nat. Commun.}\ }\textbf {\bibinfo {volume}
			{4}},\ \bibinfo {pages} {1818} (\bibinfo {year} {2013})}\BibitemShut
	{NoStop}%
	\bibitem [{\citenamefont {Luo}\ \emph {et~al.}(2015)\citenamefont {Luo},
		\citenamefont {Herrmann}, \citenamefont {Krapick}, \citenamefont {Brecht},
		\citenamefont {Ricken}, \citenamefont {Quiring}, \citenamefont {Suche},
		\citenamefont {Sohler},\ and\ \citenamefont
		{Silberhorn}}]{Luo_NewJPhys_2015}%
	\BibitemOpen
	\bibfield  {author} {\bibinfo {author} {\bibfnamefont {K.-H.}\ \bibnamefont
			{Luo}}, \bibinfo {author} {\bibfnamefont {H.}~\bibnamefont {Herrmann}},
		\bibinfo {author} {\bibfnamefont {S.}~\bibnamefont {Krapick}}, \bibinfo
		{author} {\bibfnamefont {B.}~\bibnamefont {Brecht}}, \bibinfo {author}
		{\bibfnamefont {R.}~\bibnamefont {Ricken}}, \bibinfo {author} {\bibfnamefont
			{V.}~\bibnamefont {Quiring}}, \bibinfo {author} {\bibfnamefont
			{H.}~\bibnamefont {Suche}}, \bibinfo {author} {\bibfnamefont
			{W.}~\bibnamefont {Sohler}}, and\ \bibinfo {author} {\bibfnamefont
			{C.}~\bibnamefont {Silberhorn}},\ }\bibfield  {title} {\bibinfo {title}
		{Direct generation of genuine single-longitudinal-mode narrowband photon
			pairs},\ }\href {https://doi.org/10.1088/1367-2630/17/7/073039} {\bibfield
		{journal} {\bibinfo  {journal} {New J. Phys.}\ }\textbf {\bibinfo {volume}
			{17}},\ \bibinfo {pages} {073039} (\bibinfo {year} {2015})}\BibitemShut
	{NoStop}%
	\bibitem [{\citenamefont {Ramelow}\ \emph {et~al.}(2015)\citenamefont
		{Ramelow}, \citenamefont {Farsi}, \citenamefont {Clemmen}, \citenamefont
		{Orquiza}, \citenamefont {Luke}, \citenamefont {Lipson},\ and\ \citenamefont
		{Gaeta}}]{Ramelow_ArXiv_2015}%
	\BibitemOpen
	\bibfield  {author} {\bibinfo {author} {\bibfnamefont {S.}~\bibnamefont
			{Ramelow}}, \bibinfo {author} {\bibfnamefont {A.}~\bibnamefont {Farsi}},
		\bibinfo {author} {\bibfnamefont {S.}~\bibnamefont {Clemmen}}, \bibinfo
		{author} {\bibfnamefont {D.}~\bibnamefont {Orquiza}}, \bibinfo {author}
		{\bibfnamefont {K.}~\bibnamefont {Luke}}, \bibinfo {author} {\bibfnamefont
			{M.}~\bibnamefont {Lipson}}, and\ \bibinfo {author} {\bibfnamefont {A.~L.}\
			\bibnamefont {Gaeta}},\ }\bibfield  {title} {\bibinfo {title}
		{Silicon-nitride platform for narrowband entangled photon generation},\
	}\href {http://arxiv.org/abs/1508.04358} {\bibfield  {journal} {\bibinfo
			{journal} {arXiv:1508.04358}\ } (\bibinfo {year} {2015})}\BibitemShut
	{NoStop}%
	\bibitem [{\citenamefont {He}\ \emph {et~al.}(2015)\citenamefont {He},
		\citenamefont {Bell}, \citenamefont {Casas-Bedoya}, \citenamefont {Zhang},
		\citenamefont {Clark}, \citenamefont {Xiong},\ and\ \citenamefont
		{Eggleton}}]{He_Optica_2015}%
	\BibitemOpen
	\bibfield  {author} {\bibinfo {author} {\bibfnamefont {J.}~\bibnamefont
			{He}}, \bibinfo {author} {\bibfnamefont {B.~A.}\ \bibnamefont {Bell}},
		\bibinfo {author} {\bibfnamefont {A.}~\bibnamefont {Casas-Bedoya}}, \bibinfo
		{author} {\bibfnamefont {Y.}~\bibnamefont {Zhang}}, \bibinfo {author}
		{\bibfnamefont {A.~S.}\ \bibnamefont {Clark}}, \bibinfo {author}
		{\bibfnamefont {C.}~\bibnamefont {Xiong}}, and\ \bibinfo {author}
		{\bibfnamefont {B.~J.}\ \bibnamefont {Eggleton}},\ }\bibfield  {title}
	{\bibinfo {title} {Ultracompact quantum splitter of degenerate photon
			pairs},\ }\href {https://doi.org/10.1364/OPTICA.2.000779} {\bibfield
		{journal} {\bibinfo  {journal} {Optica}\ }\textbf {\bibinfo {volume} {2}},\
		\bibinfo {pages} {779} (\bibinfo {year} {2015})}\BibitemShut {NoStop}%
	\bibitem [{\citenamefont {Vernon}\ \emph {et~al.}(2016)\citenamefont {Vernon},
		\citenamefont {Liscidini},\ and\ \citenamefont {Sipe}}]{Vernon_OL_2016}%
	\BibitemOpen
	\bibfield  {author} {\bibinfo {author} {\bibfnamefont {Z.}~\bibnamefont
			{Vernon}}, \bibinfo {author} {\bibfnamefont {M.}~\bibnamefont {Liscidini}},
		and\ \bibinfo {author} {\bibfnamefont {J.~E.}\ \bibnamefont {Sipe}},\
	}\bibfield  {title} {\bibinfo {title} {No free lunch: the trade-off between
			heralding rate and efficiency in microresonator-based heralded single photon
			sources},\ }\href {https://doi.org/10.1364/OL.41.000788} {\bibfield
		{journal} {\bibinfo  {journal} {Opt. Lett.}\ }\textbf {\bibinfo {volume}
			{41}},\ \bibinfo {pages} {788} (\bibinfo {year} {2016})}\BibitemShut
	{NoStop}%
	\bibitem [{\citenamefont {Christensen}\ \emph {et~al.}(2018)\citenamefont
		{Christensen}, \citenamefont {Koefoed}, \citenamefont {Rottwitt},\ and\
		\citenamefont {McKinstrie}}]{Christensen_OL_2018}%
	\BibitemOpen
	\bibfield  {author} {\bibinfo {author} {\bibfnamefont {J.~B.}\ \bibnamefont
			{Christensen}}, \bibinfo {author} {\bibfnamefont {J.~G.}\ \bibnamefont
			{Koefoed}}, \bibinfo {author} {\bibfnamefont {K.}~\bibnamefont {Rottwitt}},
		and\ \bibinfo {author} {\bibfnamefont {C.~J.}\ \bibnamefont {McKinstrie}},\
	}\bibfield  {title} {\bibinfo {title} {Engineering spectrally unentangled
			photon pairs from nonlinear microring resonators by pump manipulation},\
	}\href {https://doi.org/10.1364/OL.43.000859} {\bibfield  {journal} {\bibinfo
			{journal} {Opt. Lett.}\ }\textbf {\bibinfo {volume} {43}},\ \bibinfo {pages}
		{859} (\bibinfo {year} {2018})}\BibitemShut {NoStop}%
	\bibitem [{\citenamefont {Lu}\ \emph {et~al.}(2019)\citenamefont {Lu},
		\citenamefont {Li}, \citenamefont {Westly}, \citenamefont {Moille},
		\citenamefont {Singh}, \citenamefont {Anant},\ and\ \citenamefont
		{Srinivasan}}]{Lu_NatPhys_2019}%
	\BibitemOpen
	\bibfield  {author} {\bibinfo {author} {\bibfnamefont {X.}~\bibnamefont
			{Lu}}, \bibinfo {author} {\bibfnamefont {Q.}~\bibnamefont {Li}}, \bibinfo
		{author} {\bibfnamefont {D.~A.}\ \bibnamefont {Westly}}, \bibinfo {author}
		{\bibfnamefont {G.}~\bibnamefont {Moille}}, \bibinfo {author} {\bibfnamefont
			{A.}~\bibnamefont {Singh}}, \bibinfo {author} {\bibfnamefont
			{V.}~\bibnamefont {Anant}}, and\ \bibinfo {author} {\bibfnamefont
			{K.}~\bibnamefont {Srinivasan}},\ }\bibfield  {title} {\bibinfo {title}
		{Chip-integrated visible–telecom entangled photon pair source for quantum
			communication},\ }\href {https://doi.org/10.1038/s41567-018-0394-3}
	{\bibfield  {journal} {\bibinfo  {journal} {Nat. Phys.}\ }\textbf {\bibinfo
			{volume} {15}},\ \bibinfo {pages} {373} (\bibinfo {year} {2019})}\BibitemShut
	{NoStop}%
	\bibitem [{\citenamefont {Kok}\ \emph {et~al.}(2007)\citenamefont {Kok},
		\citenamefont {Munro}, \citenamefont {Nemoto}, \citenamefont {Ralph},
		\citenamefont {Dowling},\ and\ \citenamefont {Milburn}}]{Kok_RMP_2007}%
	\BibitemOpen
	\bibfield  {author} {\bibinfo {author} {\bibfnamefont {P.}~\bibnamefont
			{Kok}}, \bibinfo {author} {\bibfnamefont {W.~J.}\ \bibnamefont {Munro}},
		\bibinfo {author} {\bibfnamefont {K.}~\bibnamefont {Nemoto}}, \bibinfo
		{author} {\bibfnamefont {T.~C.}\ \bibnamefont {Ralph}}, \bibinfo {author}
		{\bibfnamefont {J.~P.}\ \bibnamefont {Dowling}}, and\ \bibinfo {author}
		{\bibfnamefont {G.~J.}\ \bibnamefont {Milburn}},\ }\bibfield  {title}
	{\bibinfo {title} {Linear optical quantum computing with photonic qubits},\
	}\href {https://doi.org/10.1103/RevModPhys.79.135} {\bibfield  {journal}
		{\bibinfo  {journal} {Rev. Mod. Phys.}\ }\textbf {\bibinfo {volume} {79}},\
		\bibinfo {pages} {135} (\bibinfo {year} {2007})}\BibitemShut {NoStop}%
	\bibitem [{\citenamefont {Ou}\ and\ \citenamefont {Lu}(1999)}]{Ou_PRL_1999}%
	\BibitemOpen
	\bibfield  {author} {\bibinfo {author} {\bibfnamefont {Z.~Y.}\ \bibnamefont
			{Ou}}and\ \bibinfo {author} {\bibfnamefont {Y.~J.}\ \bibnamefont {Lu}},\
	}\bibfield  {title} {\bibinfo {title} {Cavity enhanced spontaneous parametric
			down-conversion for the prolongation of correlation time between conjugate
			photons},\ }\href {https://doi.org/10.1103/PhysRevLett.83.2556} {\bibfield
		{journal} {\bibinfo  {journal} {Phys. Rev. Lett.}\ }\textbf {\bibinfo
			{volume} {83}},\ \bibinfo {pages} {2556} (\bibinfo {year}
		{1999})}\BibitemShut {NoStop}%
	\bibitem [{\citenamefont {Clemmen}\ \emph {et~al.}(2009)\citenamefont
		{Clemmen}, \citenamefont {Huy}, \citenamefont {Bogaerts}, \citenamefont
		{Baets}, \citenamefont {Emplit},\ and\ \citenamefont
		{Massar}}]{Clemmen_OE_2009}%
	\BibitemOpen
	\bibfield  {author} {\bibinfo {author} {\bibfnamefont {S.}~\bibnamefont
			{Clemmen}}, \bibinfo {author} {\bibfnamefont {K.~P.}\ \bibnamefont {Huy}},
		\bibinfo {author} {\bibfnamefont {W.}~\bibnamefont {Bogaerts}}, \bibinfo
		{author} {\bibfnamefont {R.~G.}\ \bibnamefont {Baets}}, \bibinfo {author}
		{\bibfnamefont {P.}~\bibnamefont {Emplit}}, and\ \bibinfo {author}
		{\bibfnamefont {S.}~\bibnamefont {Massar}},\ }\bibfield  {title} {\bibinfo
		{title} {Continuous wave photon pair generation in silicon-on-insulator
			waveguides and ring resonators},\ }\href
	{https://doi.org/10.1364/OE.17.016558} {\bibfield  {journal} {\bibinfo
			{journal} {Opt. Express}\ }\textbf {\bibinfo {volume} {17}},\ \bibinfo
		{pages} {16558} (\bibinfo {year} {2009})}\BibitemShut {NoStop}%
	\bibitem [{\citenamefont {Ell}\ \emph {et~al.}(2001)\citenamefont {Ell},
		\citenamefont {Morgner}, \citenamefont {K\"{a}rtner}, \citenamefont
		{Fujimoto}, \citenamefont {Ippen}, \citenamefont {Scheuer}, \citenamefont
		{Angelow}, \citenamefont {Tschudi}, \citenamefont {Lederer}, \citenamefont
		{Boiko},\ and\ \citenamefont {Luther-Davies}}]{Ell_OL_2001}%
	\BibitemOpen
	\bibfield  {author} {\bibinfo {author} {\bibfnamefont {R.}~\bibnamefont
			{Ell}}, \bibinfo {author} {\bibfnamefont {U.}~\bibnamefont {Morgner}},
		\bibinfo {author} {\bibfnamefont {F.~X.}\ \bibnamefont {K\"{a}rtner}},
		\bibinfo {author} {\bibfnamefont {J.~G.}\ \bibnamefont {Fujimoto}}, \bibinfo
		{author} {\bibfnamefont {E.~P.}\ \bibnamefont {Ippen}}, \bibinfo {author}
		{\bibfnamefont {V.}~\bibnamefont {Scheuer}}, \bibinfo {author} {\bibfnamefont
			{G.}~\bibnamefont {Angelow}}, \bibinfo {author} {\bibfnamefont
			{T.}~\bibnamefont {Tschudi}}, \bibinfo {author} {\bibfnamefont {M.~J.}\
			\bibnamefont {Lederer}}, \bibinfo {author} {\bibfnamefont {A.}~\bibnamefont
			{Boiko}}, and\ \bibinfo {author} {\bibfnamefont {B.}~\bibnamefont
			{Luther-Davies}},\ }\bibfield  {title} {\bibinfo {title} {Generation of 5-fs
			pulses and octave-spanning spectra directly from a {T}i:sapphire laser},\
	}\href {https://doi.org/10.1364/OL.26.000373} {\bibfield  {journal} {\bibinfo
			{journal} {Opt. Lett.}\ }\textbf {\bibinfo {volume} {26}},\ \bibinfo {pages}
		{373} (\bibinfo {year} {2001})}\BibitemShut {NoStop}%
	\bibitem [{\citenamefont {Coen}\ \emph {et~al.}(2013)\citenamefont {Coen},
		\citenamefont {Randle}, \citenamefont {Sylvestre},\ and\ \citenamefont
		{Erkintalo}}]{Coen_OL_2013}%
	\BibitemOpen
	\bibfield  {author} {\bibinfo {author} {\bibfnamefont {S.}~\bibnamefont
			{Coen}}, \bibinfo {author} {\bibfnamefont {H.~G.}\ \bibnamefont {Randle}},
		\bibinfo {author} {\bibfnamefont {T.}~\bibnamefont {Sylvestre}}, and\
		\bibinfo {author} {\bibfnamefont {M.}~\bibnamefont {Erkintalo}},\ }\bibfield
	{title} {\bibinfo {title} {Modeling of octave-spanning kerr frequency combs
			using a generalized mean-field lugiato-lefever model},\ }\href
	{https://doi.org/10.1364/OL.38.000037} {\bibfield  {journal} {\bibinfo
			{journal} {Opt. Lett.}\ }\textbf {\bibinfo {volume} {38}},\ \bibinfo {pages}
		{37} (\bibinfo {year} {2013})}\BibitemShut {NoStop}%
	\bibitem [{\citenamefont {Fortier}\ \emph {et~al.}(2006)\citenamefont
		{Fortier}, \citenamefont {Bartels},\ and\ \citenamefont
		{Diddams}}]{Fortier_OL_2006}%
	\BibitemOpen
	\bibfield  {author} {\bibinfo {author} {\bibfnamefont {T.~M.}\ \bibnamefont
			{Fortier}}, \bibinfo {author} {\bibfnamefont {A.}~\bibnamefont {Bartels}},
		and\ \bibinfo {author} {\bibfnamefont {S.~A.}\ \bibnamefont {Diddams}},\
	}\bibfield  {title} {\bibinfo {title} {Octave-spanning {T}i:sapphire laser
			with a repetition rate \textgreater 1 {GH}z for optical frequency
			measurements and comparisons},\ }\href {https://doi.org/10.1364/OL.31.001011}
	{\bibfield  {journal} {\bibinfo  {journal} {Opt. Lett.}\ }\textbf {\bibinfo
			{volume} {31}},\ \bibinfo {pages} {1011} (\bibinfo {year}
		{2006})}\BibitemShut {NoStop}%
\end{thebibliography}

\begin{thebibliography}{3}%
	\makeatletter
	\providecommand \@ifxundefined [1]{%
		\@ifx{#1\undefined}
	}%
	\providecommand \@ifnum [1]{%
		\ifnum #1\expandafter \@firstoftwo
		\else \expandafter \@secondoftwo
		\fi
	}%
	\providecommand \@ifx [1]{%
		\ifx #1\expandafter \@firstoftwo
		\else \expandafter \@secondoftwo
		\fi
	}%
	\providecommand \natexlab [1]{#1}%
	\providecommand \enquote  [1]{``#1''}%
	\providecommand \bibnamefont  [1]{#1}%
	\providecommand \bibfnamefont [1]{#1}%
	\providecommand \citenamefont [1]{#1}%
	\providecommand \href@noop [0]{\@secondoftwo}%
	\providecommand \href [0]{\begingroup \@sanitize@url \@href}%
	\providecommand \@href[1]{\@@startlink{#1}\@@href}%
	\providecommand \@@href[1]{\endgroup#1\@@endlink}%
	\providecommand \@sanitize@url [0]{\catcode `\\12\catcode `\$12\catcode
		`\&12\catcode `\#12\catcode `\^12\catcode `\_12\catcode `\%12\relax}%
	\providecommand \@@startlink[1]{}%
	\providecommand \@@endlink[0]{}%
	\providecommand \url  [0]{\begingroup\@sanitize@url \@url }%
	\providecommand \@url [1]{\endgroup\@href {#1}{\urlprefix }}%
	\providecommand \urlprefix  [0]{URL }%
	\providecommand \Eprint [0]{\href }%
	\providecommand \doibase [0]{https://doi.org/}%
	\providecommand \selectlanguage [0]{\@gobble}%
	\providecommand \bibinfo  [0]{\@secondoftwo}%
	\providecommand \bibfield  [0]{\@secondoftwo}%
	\providecommand \translation [1]{[#1]}%
	\providecommand \BibitemOpen [0]{}%
	\providecommand \bibitemStop [0]{}%
	\providecommand \bibitemNoStop [0]{.\EOS\space}%
	\providecommand \EOS [0]{\spacefactor3000\relax}%
	\providecommand \BibitemShut  [1]{\csname bibitem#1\endcsname}%
	\let\auto@bib@innerbib\@empty
	\bibitem [{\citenamefont {Joshi}\ \emph {et~al.}(2016)\citenamefont {Joshi},
		\citenamefont {Jang}, \citenamefont {Luke}, \citenamefont {Ji}, \citenamefont
		{Miller}, \citenamefont {Klenner}, \citenamefont {Okawachi}, \citenamefont
		{Lipson},\ and\ \citenamefont {Gaeta}}]{Joshi_OL_2016}%
	\BibitemOpen
	\bibfield  {author} {\bibinfo {author} {\bibfnamefont {C.}~\bibnamefont
			{Joshi}}, \bibinfo {author} {\bibfnamefont {J.~K.}\ \bibnamefont {Jang}},
		\bibinfo {author} {\bibfnamefont {K.}~\bibnamefont {Luke}}, \bibinfo {author}
		{\bibfnamefont {X.}~\bibnamefont {Ji}}, \bibinfo {author} {\bibfnamefont
			{S.~A.}\ \bibnamefont {Miller}}, \bibinfo {author} {\bibfnamefont
			{A.}~\bibnamefont {Klenner}}, \bibinfo {author} {\bibfnamefont
			{Y.}~\bibnamefont {Okawachi}}, \bibinfo {author} {\bibfnamefont
			{M.}~\bibnamefont {Lipson}},\ and\ \bibinfo {author} {\bibfnamefont {A.~L.}\
			\bibnamefont {Gaeta}},\ }\bibfield  {title} {\bibinfo {title} {Thermally
			controlled comb generation and soliton modelocking in microresonators},\
	}\href {https://doi.org/10.1364/OL.41.002565} {\bibfield  {journal} {\bibinfo
			{journal} {Opt. Lett.}\ }\textbf {\bibinfo {volume} {41}},\ \bibinfo {pages}
		{2565} (\bibinfo {year} {2016})}\BibitemShut {NoStop}%
	\bibitem [{\citenamefont {Ou}\ and\ \citenamefont {Lu}(1999)}]{Ou_PRL_1999}%
	\BibitemOpen
	\bibfield  {author} {\bibinfo {author} {\bibfnamefont {Z.~Y.}\ \bibnamefont
			{Ou}}\ and\ \bibinfo {author} {\bibfnamefont {Y.~J.}\ \bibnamefont {Lu}},\
	}\bibfield  {title} {\bibinfo {title} {Cavity enhanced spontaneous parametric
			down-conversion for the prolongation of correlation time between conjugate
			photons},\ }\href {https://doi.org/10.1103/PhysRevLett.83.2556} {\bibfield
		{journal} {\bibinfo  {journal} {Phys. Rev. Lett.}\ }\textbf {\bibinfo
			{volume} {83}},\ \bibinfo {pages} {2556} (\bibinfo {year}
		{1999})}\BibitemShut {NoStop}%
	\bibitem [{\citenamefont {Gardiner}\ and\ \citenamefont
		{Collett}(1985)}]{Gardiner_PRA_1985}%
	\BibitemOpen
	\bibfield  {author} {\bibinfo {author} {\bibfnamefont {C.~W.}\ \bibnamefont
			{Gardiner}}\ and\ \bibinfo {author} {\bibfnamefont {M.~J.}\ \bibnamefont
			{Collett}},\ }\bibfield  {title} {\bibinfo {title} {Input and output in
			damped quantum systems: Quantum stochastic differential equations and the
			master equation},\ }\href {https://doi.org/10.1103/PhysRevA.31.3761}
	{\bibfield  {journal} {\bibinfo  {journal} {Phys. Rev. A}\ }\textbf {\bibinfo
			{volume} {31}},\ \bibinfo {pages} {3761} (\bibinfo {year}
		{1985})}\BibitemShut {NoStop}%
\end{thebibliography}
\end{document}